\documentclass[aps,prd,twocolumn,superscriptaddress,floatfix]{revtex4-2}
\usepackage[pdftex]{graphicx}

\usepackage{xcolor}
\usepackage{float}
\usepackage{amsmath,amssymb,amsfonts,amsbsy}
\usepackage{hyperref}  

\usepackage[utf8]{inputenc}

\newcommand{\pbar}{{\bar{p}}}

\newcommand{\bk}{{\bf k}}
\newcommand{\bK}{{\bf K}}
\newcommand{\bP}{{\bf P}}
\newcommand{\cM}{{\mathcal{M}}}
\newcommand{\cJ}{{\mathcal{J}}}
\newcommand{\bphi}{{\bar{\phi}}}
\newcommand{\GeV}{\mathrm{~GeV}}
\newcommand{\MeV}{\mathrm{~MeV}}
\renewcommand{\Re}{\mathrm{Re }}
\renewcommand{\Im}{\mathrm{Im }}

\begin{document}
	

\title{Studying time-like proton form factors using vortex state \texorpdfstring{$p\bar{p}$}{ppBar} annihilation}


\author{Nikolai Korchagin}
\email{nikolaik@mail.sysu.edu.cn}
\affiliation{School of Physics and Astronomy, Sun Yat-sen University, Zhuhai 519082, China}


\begin{abstract}
Vortex states of particles --- non-plane-wave solutions of the corresponding wave equation with a helicoidal wave front --- open new opportunities for particle physics, unavailable in plane wave scattering.
Here we demonstrate that $p\bar p$ annihilation with a vortex proton and antiproton provides access to the phase of proton electromagnetic form factors even in unpolarized scattering.
\end{abstract}

\pacs{}

\maketitle


\section{Introduction}

A twisted (or vortex) state of a field is a non–plane-wave solution of a wave equation that, on average, propagates along a certain direction (e.g., axis z) and possesses a non-zero z-projection of the intrinsic orbital angular momentum (OAM). 

Vortex states of optical photons were produced in 1990s and have since found numerous applications 
\cite{Torres2011:applicationsOAMlight,andrews2012angular,Allen:1999,Molina-Terriza:2007,Padgett2017:OAM25years,Knyazev-Serbo:2018,babiker2018:atomsInTwLight}. Twisted electrons were suggested in \cite{Bliokh2007:semiclassical}, and shortly afterwards were experimentally produced~\cite{UchidaTonomura:Twe,Verbeeck2010:Twe,McMorran:2011}. 
The energy of these vortex electrons was low, up to 300 keV.
There are proposals to put high-energy electrons in vortex states or to accelerate low-energy vortex electrons to higher energies, but none of them has been realized yet. 
Moreover, cold neutrons~\cite{clark2015:TwN,sarenac2018methods,sarenac2019generation,Sarenac:2022} and even atoms~\cite{luski2021atoms} in twisted states have been achieved experimentally.

In the last 10 years, the application of twisted states has become a flourishing research field in optics \cite{Torres2011:applicationsOAMlight,andrews2012angular,Allen:1999,Molina-Terriza:2007,Padgett2017:OAM25years,Knyazev-Serbo:2018,babiker2018:atomsInTwLight}, microscopy \cite{Verbeeck2010:Twe,Juchtmans2015,Juchtmans2015a}, quantum communication \cite{Leach2010,Wang2012}, atomic and condensed matter physics. 
However, in the particle and nuclear physics communities, vortex states are not well known, mainly because experiments reported so far are limited to very low energies.

Nevertheless, interest is growing, and there have been several proposals on how to bring vortex states of photons and massive particles to energies in the MeV or GeV range \cite{Karlovets2022a,Karlovets2022}. When demonstrated experimentally, they will offer new opportunities for nuclear, hadron physics, quantum electrodynamics.

Creating high-energetic twisted particles requires novel experimental setups and dedicated efforts. Nevertheless, certain collision experiments with twisted photons, electrons, and nucleons can be performed using existing technology.

The potential of vortex states for application in high-energy physics was demonstrated in number of cases. Some of them are:
\begin{itemize}
	\item scattering of two vortex states allows one to probe the overall phase of the scattering amplitude, giving direct experimental access, for instance, to the Coulomb phase \cite{Ivanov2016:CoulombPhase,Ivanov2012:phase,Karlovets2016,Karlovets2017};
	\item vortex states could be used as a new tool for spin physics allowing to probe spin and parity-dependent observables with fully inclusive unpolarized cross sections \cite{Ivanov:2019prl,Ivanov:2019kinematics,Ivanov:2020newtool};
	\item in decay of vortex muon, the energy spectrum of electron is a sensitive probe of the vortex muon properties \cite{Zhao2021:muonDecay,Zhao2023:muonDecay2};
	\item in nuclear excitation by electron capture, if the electron energy matches the resonance condition, one can excite a nucleus to
	an isomeric state. It was found that using electron in vortex state could significantly modify the decay rate of the isomers \cite{Gargiulo2022,Wu2022};
	\item in nuclear physics the giant resonances with specific multipolarity can be extracted via vortex $\gamma$-photons \cite{Lu2023}.
\end{itemize} 
More examples could be found in reviews \cite{Ivanov2022review,Bliokh2017}.

Although twisted protons or other hadrons have not been demonstrated yet, once they are, they will give a new playground for particle physics, enabling to study numerous phenomena even with non-relativistic protons. 
For example, due to proton's high mass, the energy released in $p\bar p$ annihilation at rest is sufficient to produce strongly interacting particles, such as pions and kaons, as well as electromagnetically interacting ones (photons, electrons and muons).

The goals of this work is to demonstrate yet another possible application of vortex states in particle physics, specifically, how it can be used to study electromagnetic form factors (FFs) of the proton.
This paper shows that in twisted $p\bar p$ annihilation to $e^+ e^-$, it is possible to probe the phase of FFs even when particles are unpolarized. 

Throughout the paper three-dimensional vectors will be denoted by symbols with arrows, and the transverse momenta will be by bold.


\section{Twisted states and their scattering}\label{section:intro_twisted_states}
This section gives a qualitative picture of vortex states and properties of double-vortex scattering. 
More detailed description can be found in Appendices~\ref{secA:twisted_fermion}, \ref{secA:twScattering} and reviews \cite{Ivanov2022review,Bliokh2017}.

What distinguishes a vortex from a plane wave (PW) is the structure of the wave front.
A plane wave has flat phase fronts that are globally normal to the propagation direction. 
In contrast, the phase front of the vortex wave has a helical shape winding around the axis of propagation. As a result, the phase depends on the angular position about the axis.

The simplest type of a vortex state is the so-called Bessel state. 
In cylindrical coordinates $(\rho,\phi_r,z)$, a scalar particle with mass $m$, propagating on average along $z$ direction, is characterized by a function
\begin{equation}\label{eq:Bessel_idea}
\begin{split}
    \Psi(r,t) &= \frac{N}{\sqrt{2E}} \psi_{l, \varkappa}(\mathbf{r})  e^{-i E t+ik_z z} , \\
    \psi_{l,\varkappa}(\mathbf{r}) &= \sqrt{\frac{\varkappa}{2\pi} } J_l(\rho \varkappa) e^{i l \phi_r},
\end{split}    
\end{equation}
where $E=\sqrt{\vec{k}^2+m^2}$ and $N$ is the normalization factor.
The integer topological charge $l$ (also called the winding number) in the radial part quantizes the winding, such that the phase changes $2\pi l$ during a full rotation about the axis. The phase factor of $e^{il\phi_r}$, that gives rise to this helical phase structure, is a characteristic feature of orbital angular momentum. For example, a similar phase factor appears in the azimuthal components of the electron wave functions in the hydrogen atom. 

\begin{figure}
	\centering
	\includegraphics[width=0.48\linewidth]{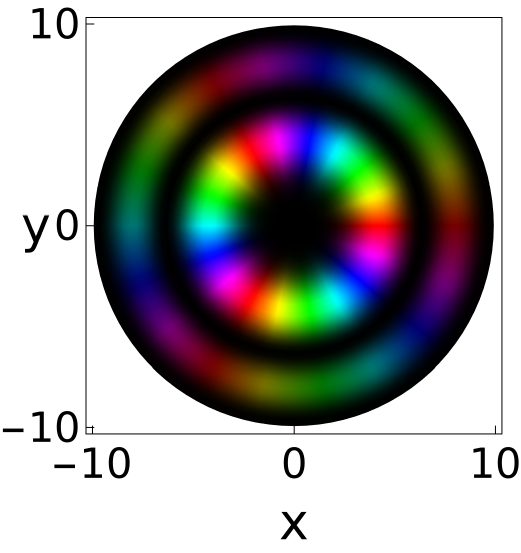}
	\includegraphics[width=0.48\linewidth]{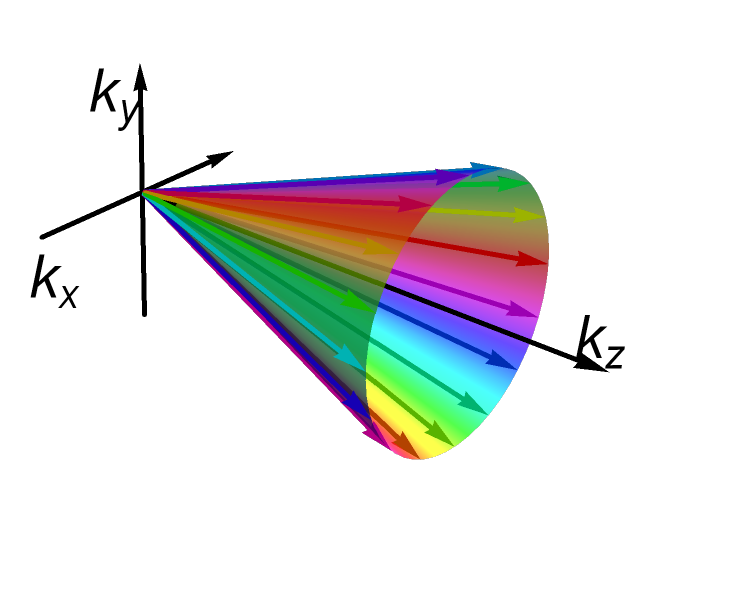}
	\caption{The Bessel state of a scalar particle in the coordinate space (left) and in momentum space (right) with $l=3$. The intensity of color indicates the probability and color itself encodes the phase. In momentum space every plane wave component has its own phase and azimuthal direction.}
	\label{fig:bessel-illustration}
\end{figure}

The helical phase structure of the vortex beam leads to an indeterminate phase at the axis of the beam, since it is connected to all possible phases of the wave. This central phase singularity is compensated by vanishing of $\psi_{l,\varkappa}(\mathbf{r})$ on axis (at the location of the singularity). For $l \neq 0$, radial part $\psi(0)=0$ in \eqref{eq:Bessel_idea} so that the intensity is exactly zero on the axis. This gives the beam a cross-sectional distribution in the form of a ring, Fig.~\ref{fig:bessel-illustration}.

In the momentum space, the Bessel state forms a circle of radius $\varkappa \equiv |\mathbf{k}|$ with offset $k_z$ from the origin along $z$ axis, Fig.~\ref{fig:bessel-illustration}. 
The radial part of the wave function $\psi_{l,\varkappa}(\mathbf{r})$ is related to its momentum representation $a_{l,\varkappa}(\mathbf{k})$ by the usual Fourier transform  \cite{JentschuraSerbo2011,JentschuraSerbo:ComptonUpconversion},
\begin{equation}
\begin{split}
    \psi_{l,\varkappa}(\mathbf{r}) &= \int \frac{d^2\mathbf{k}}{(2\pi)^2} a_{l,\varkappa}(\mathbf{k}) e^{i \mathbf{k} \cdot \mathbf{r}} , 
    \\
    a_{l,\varkappa}(\mathbf{k}) &= {(-i)}^l e^{i l \phi_k} \sqrt{\frac{2\pi}{\varkappa}} \delta(|\mathbf{k}|-\varkappa).
\end{split}
\end{equation}
$\varkappa$ quantifies the transverse momentum content, hidden inside the Bessel state, although on average $<\!\! \mathbf{k} \!\!>=0$.
The momentum vectors of the plane wave components form a cone with the opening angle $\theta$.  
The paraxial regime, which will often be used below, corresponds to $\theta \ll 1$. 
Other vortex states, like the Laguerre-Gaussian state, can be regarded as superpositions of many Bessel type vortex states. 
In this paper we work only with pure Bessel states, because it simplifies the  calculation for scattering amplitudes.

Until now, we have discussed scalar particles. It is possible to construct a vortex state for a particle with an internal spin degree of freedom, like electron or photon (see~\cite{Ivanov2022review}). 
We give a detailed description of spin-$1/2$ fermion in Bessel state in Appendix~\ref{secA:twisted_fermion}.

Now let us consider the scattering of two Bessel states.
For simplicity, now we will focus on spinless particles in order to demonstrate novel kinematical effects, inherent for double-vortex scattering.
We define two counter-propagating Bessel states with a common axis $z$. 
The final state is represented by two particles described as plane waves.

Bessel states in transverse momentum space form circles of radius $\varkappa_1$ and $\varkappa_2$. 
$\mathbf{k}_{1}$ and $\mathbf{k}_{2}$ sum up to the total transverse momenta $\mathbf{K} = \mathbf{k}_{1}+\mathbf{k}_{2}$. Since angles $\phi_1$ and $\phi_2$ are arbitrary, $\mathbf{K}$ spans a ring (possible values of $\mathbf{K}$ will lay inside the ring $|\varkappa_1-\varkappa_2|<\mathbf{K} <\varkappa_1+\varkappa_2$)

It is in contrast with PW scattering, where initial states in the momentum transverse plane are points. Thus, $\bK$ is also completely fixed ($\bK=0$ in c.m.\ frame).
Therefore, the cross section in double-Bessel scattering has an additional distribution over the total transverse momenta $\bK$.

\begin{figure}
	\centering
	\includegraphics[width=0.95\linewidth]{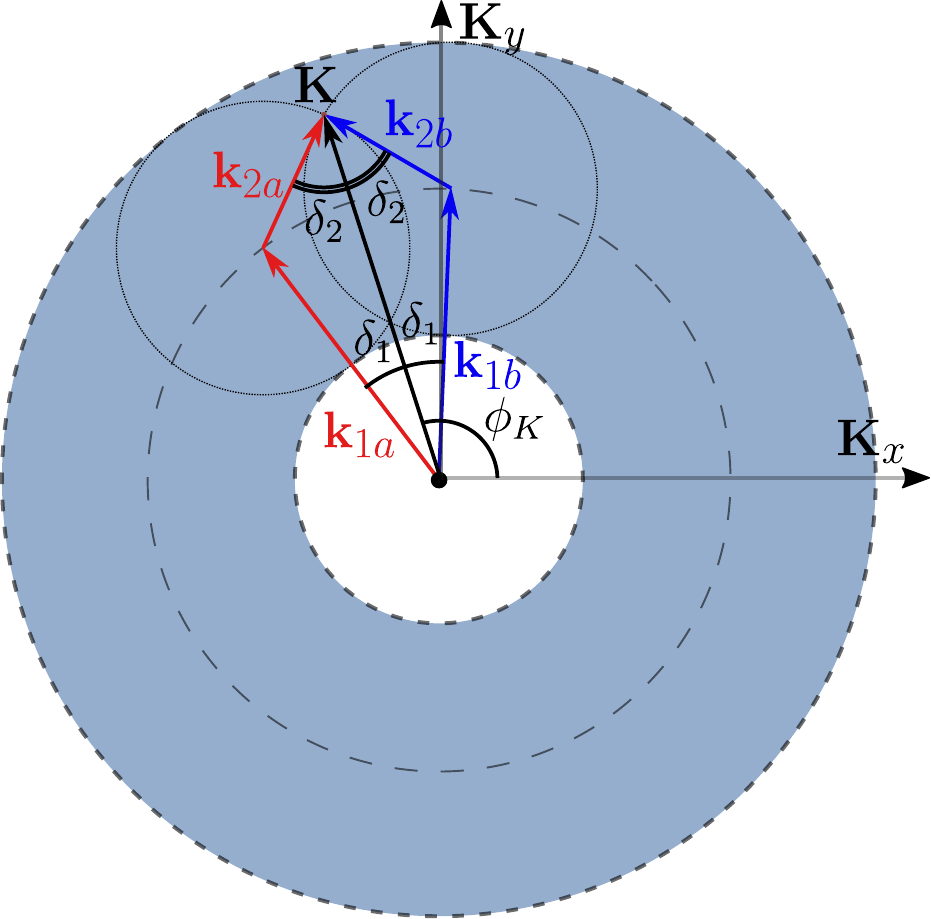}
	\caption{The two plane wave configurations in the transverse momentum plane that can lead to the final transverse momentum $\bK$ in double-Bessel scattering. The shaded ring represents possible values for $\bK$}
	\label{fig:2momenta_conf}
\end{figure}

For final particles, the situation is similar to usual PW scattering. Since they are described by PW states, their transverse momenta sum up to some fixed $\mathbf{K}$. 
It means that a given configuration of detected final particles corresponds to some point on $\mathbf{K}$-ring. 
Although two Bessel state collision realizes many $\mathbf{K}$, for given momenta of detected particles, only one particular vector $\bK$ is selected.
There are two ways to put together the same vector $\bK$.
Any given value of $\bK$ inside the ring could be achieved by two possible combinations of vectors $\bk_{1}$ and $\bk_{2}$, Fig.~\ref{fig:2momenta_conf}.
These two possibilities, denoted $a$ and $b$, are related by reflection of $\bk_{1,2}$ with respect to $\bK$.
It is a distinguishing feature of two-vortex scattering, which leads to the fact that the total scattering amplitude is a sum of two plane wave scattering amplitudes, $\cM_a$ and $\cM_b$, with different kinematics of initial particles,
\begin{align}\label{eq:tw_amplitude_def_2}
	\mathcal{J} =& \frac{e^{i(l_1-l_2)\phi_K} }{\sin (\delta_1+ \delta_2)} \nonumber
	\\
	&\Big(
	\mathcal{M}_a e^{i(l_1\delta_1+l_2\delta_2)}	
    + 
	\mathcal{M}_b e^{-i(l_1\delta_1+l_2\delta_2)}
	\Big).
\end{align}
Due to interference between $\cM_a$ and $\cM_b$, the differential cross section demonstrates oscillations in $\mathbf{K}$ distribution, forming concentric interference fringes.
Contrast and form of fringes are defined by details of amplitudes $\mathcal{M}$ and parameters of Bessel states ($E_i$, $\varkappa_i$, $l_i$).

What will one see in such a double-vortex into two PWs scattering experiment? 
In every scattering event, provided that both final particles are detected,
one can sum final momenta and make a differential cross section distribution over $\mathbf{K}$-ring.
In current work we show that this distribution can be used to study electromagnetic FFs of nucleon.

\section{Nucleon electromagnetic form factors in time-like region}\label{section:FFs}

Understanding the internal structure of the nucleon is a cornerstone of the theory of strong interaction. 
As a composite fermion, the proton is described by two FFs:  electric $G_E(q^2)$ and magnetic $G_M(q^2)$. 
Proton FFs in space-like regime $(q^2 < 0)$ could be measured in the elastic electron-proton scattering.  The annihilation channel, in $e^+ e^- \to p\pbar$ and $p\pbar \to e^+ e^-$ reactions, is used for studying the time-like region, since $q^2>0$.
In the space-like regime, FFs are real valued functions and related to distribution of electric charge and magnetization (note that such interpretation is valid only in the Breit frame~\cite{pacetti2014:ProtonElectromagneticFF}).
In contrast, in the time-like region, FFs are complex valued functions of total energy and related through a Fourier transform in time to the transition amplitude from a virtual-photon state to a hadronic state.

In the past, independent measurement of time-like FFs was unavailable due to limited luminosities of $e^+e^-$ colliders. In the last decade, experimental efforts allowed measurement of the differential cross section and separation of $G_E$ and $G_M$.
 
In the one photon exchange approximation, the differential cross section in c.m.s. for $p \bar p \to e^+ e^-$ is
\begin{equation}\label{eq:PW-dSigma}
	\frac{d\sigma_{p\bar p \to e^+ e^-}}{d \cos \theta} = \frac{\pi \alpha^2 }{2 q^2 \beta}
	\Big[
	|G_M|^2 (1+ \cos^2 \theta ) + \frac{1}{\tau} |G_E|^2 \sin^2 \theta
	\Big],
\end{equation}
where $\beta=\sqrt{1-1/\tau}$, with $\tau=q^2/(4M^2)\geq 1$, is the velocity of the incoming proton. $\alpha$ is the fine structure constant. In annihilation channel $q^2=s$, where $s$ is the standard Mandelstam kinematical variable.
Note that Eq.~\eqref{eq:PW-dSigma} depends on the modulus of FFs. 

On the other hand, as was mentioned before, time-like FFs are complex valued functions and, therefore, have phases.
The phase of an individual FF is not observable. However, the phase difference between two FFs could be measured.
Knowledge of this phase will bring new insights about the structure of the proton. 
However, all unpolarized experiments are insensitive to the FF phase.

To measure the relative FFs phase, one can perform an experiment with polarized initial particles or measure polarizations in the final state.
If we consider the $p\pbar$ annihilation in a reference system with the $z$ axis along the antiproton beam momentum, and $xz$ is the scattering plane, the dependence of the cross section on the polarizations $P_1$ and $P_2$ of the colliding antiproton and proton can be written as \cite{pacetti2014:ProtonElectromagneticFF}
\begin{multline}
	\left(\frac{d \sigma}{d \Omega}\right)\left(\boldsymbol{P}_1, \boldsymbol{P}_2\right)=\left(\frac{d \sigma}{d \Omega}\right)_0 \Big[1+A_y\left(P_{1 y}+P_{2 y}\right)
	\\
	+A_{x x} P_{1 x} P_{2 x}+A_{y y} P_{1 y} P_{2 y}+A_{z z} P_{1 z} P_{2 z}
	\\
	+A_{x z}\left(P_{1 x} P_{2 z}+P_{1 z} P_{2 x}\right)\Big]
\end{multline}
where the coefficients $A_i$ and $A_{ij}=A_{ji}$ are analyzing powers and symmetric correlation coefficients.
Two of them are sensitive to the relative FF phase shift:
\begin{align}
\left(\frac{d\sigma}{d \Omega}\right)_0 A_y &= \frac{\pi \alpha^2}{2 \beta q^2 \sqrt{\tau}} \sin(2\theta) \Im\{G_M G_E^* \},
\\
\left(\frac{d\sigma}{d \Omega}\right)_0 A_{xz} &= \frac{\pi \alpha^2}{2 \beta q^2 \sqrt{\tau}} \sin(2\theta) \Re\{G_M G_E^* \}.
\end{align}
$A_y$ could be measured as a transverse single spin asymmetry of the cross section.  
One could also consider double-spin observables \cite{TomasiGustafsson2005} in order to extract $\Re\{G_E G_M^*\}$, which is useful if the relative FF phase is small.

Alternatively, one can measure polarization of final proton in unpolarized $e^+e^-$ annihilation\cite{Dubnickova:1992ii}.
Experimentally, it could be achieved by including a polarimeter in the experimental setup of a collider experiment.

In this work we propose an alternative method to study the relative FF phase by using vortex states, which allows us to probe the phase even in an unpolarized process.

\section{Plane wave scattering amplitudes in general kinematics}
\begin{figure}
	\centering
	\includegraphics[width=0.7\linewidth]{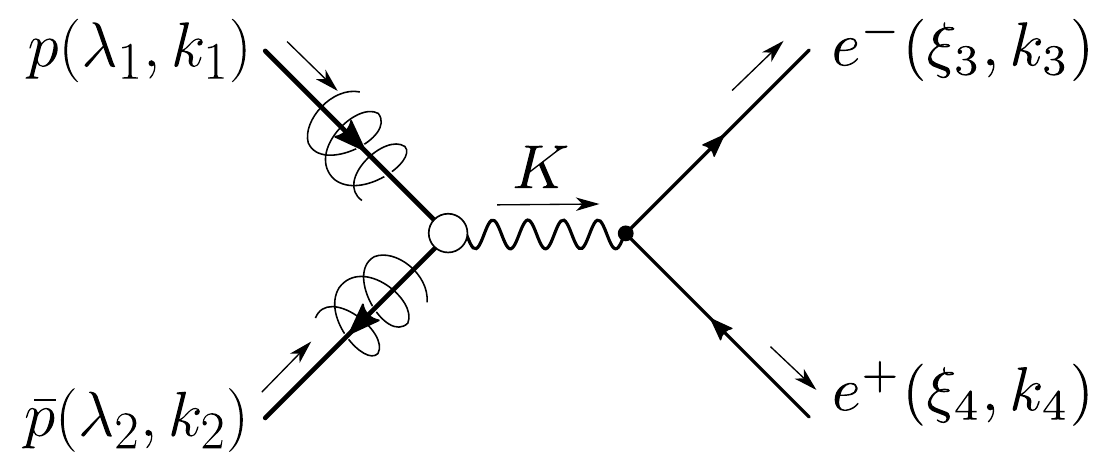}
	\caption{The Feynman diagram for the plane wave scattering amplitude. Spirals on hadron lines represent that these plane waves are components of twisted wave packets. $\lambda_i$ and $\xi_i$ stands for the nucleon and lepton helicities.}
	\label{fig:ppbartoee}
\end{figure}

We need to calculate scattering amplitude $\mathcal{M}_{a,b}$ in Eq.~\eqref{eq:tw_amplitude_def_2} keeping momenta of all participating particles arbitrary.
Consider $p(k_1) + \bar{p}(k_2) \to e^-(k_3)+e^+(k_4)$ process, Fig.~\ref{fig:ppbartoee}, with momenta denoted as:
\begin{equation}
\begin{split}
    \textrm{proton:  } &k_1=(E_1,\bk_{1} , k_{1z}), \\
    \textrm{anti-proton:  } & k_2 = (E_2,\bk_{2}, k_{2z}), \\
    \textrm{electron:  } &k_3 = (E_3, \bk_{3},k_{3z}), \\
    \textrm{positron:  } &k_4 = (E_4, \bk_{4},k_{4z}).
\end{split}
\end{equation}
Modulus of a transverse momenta will be denoted by $\varkappa$:
\begin{equation}
\begin{split}
    \bk_{i} &=\varkappa_i (\cos \phi_i ; \sin \phi_i),
	\\
    |\bk_{i}| &\equiv \varkappa_i,
	\\
	k_{z,i} &= |\vec{k}_i| \cos \theta_i.
\end{split}  
\end{equation}
Leptons are treated as massless and nucleon has mass $M$.
In the one photon exchange approximation the matrix element of the PW scattering amplitude is a product of hadron $J^\mu$ and lepton $L_\mu$ currents,
\begin{equation}
	\mathcal{M} = \frac{e^2}{s} J^\mu L_\mu,
\end{equation}
where $e$ is the electric charge in units of electron charge.

The current for the nucleon is
\begin{equation}
\begin{aligned}  \label{eq:hadron_current}
    J^{\mu} &=
	\bar{v}_{\lambda_2}(k_2)
	\left[\gamma^\mu F_1 +  \frac{F_2 }{2M} \sigma^{\mu\nu} K_{\nu} 
	\right] u_{\lambda_1}(k_1)  \\
	&= 
	\bar{v}_{\lambda_2}(k_2)
	\Bigg[ 
	\gamma^\mu G_M  
	+ \frac{P^{\mu}}{2M} \frac{G_M -G_E }{1-\tau}
	\Bigg] 
	u_{\lambda_1}(k_1),
\end{aligned}
\end{equation}
where $\sigma_{\mu\nu}=[\gamma_\mu,\gamma_\nu ]/2$, $K = k_1 + k_2$, $s=K^2$, and $P= k_2- k_1$.
To get the second line we used Gordon identity and Sach FFs: 
\begin{equation}
\begin{aligned}
	G_E=& F_1+\tau F_2,
	\\
	G_M=& F_1+F_2.
\end{aligned}
\end{equation}
\begin{figure}
	\centering
	\includegraphics[width=0.9\linewidth]{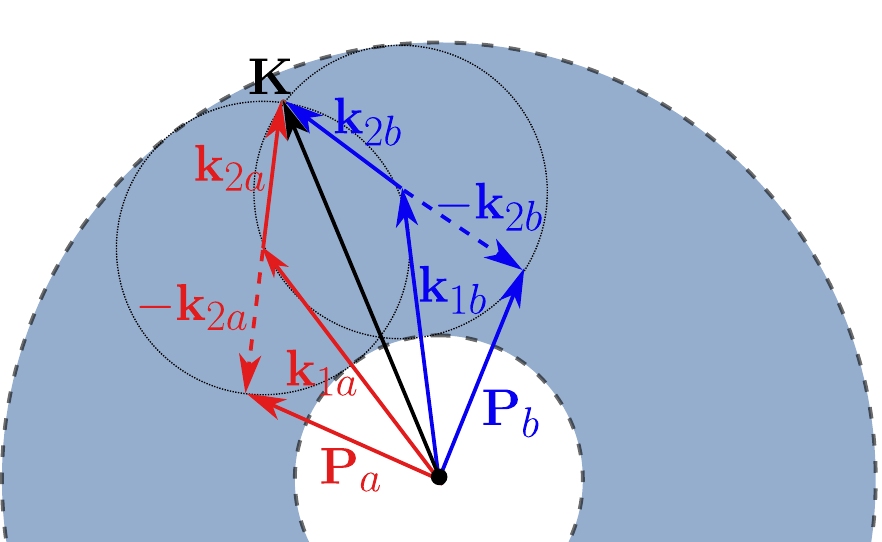}
    \caption{The transverse vector $\bP$ is different for the two plane wave amplitudes interfering in Eq.~\eqref{eq:tw_amplitude_def_2}. As a  consequence, the cross section distribution becomes sensitive to the relative form factor phase.}
	\label{fig:Pvector}
\end{figure}

As we will see further, $P^\mu$ in Eq.~\eqref{eq:hadron_current} is the reason of sensitivity of double-vortex scattering to the relative FF phase. For two plane wave configurations $\mathcal{M}_{a,b}$ in Eq.~\eqref{eq:tw_amplitude_def_2}, $\bP$ is different, Fig.~\ref{fig:Pvector}.
It is an example of the effect which was discussed before in more general settings in \cite{Karlovets2017,Karlovets2016}.

The lepton current is given by the standard formula
\begin{align}\label{eq:lepton_current}
	L^\mu = \bar{u}_{\xi_3}(k_3)\gamma^\mu  v_{\xi_4}(k_4).
\end{align}

Since we have to deal with spinors in general kinematics, the resulting equations are large and opaque to establish physical intuition. Therefore, we do full calculations numerically. However, in order to gain some insights, we also made analytical calculations in a paraxial limit, when $\varkappa_{1,2} \ll |k_{z 1,2}|$.

One can write a double-vortex differential cross section distribution $\varsigma \equiv \frac{d\sigma}{d^2\bk_{3} d^2 \mathbf{K}}$ in general form as a sum
\begin{equation*}
	\varsigma = \varsigma_0 + \varsigma_1 \sin (\phi_K-\phi_3) + \varsigma_2 \cos (\phi_K-\phi_3).
\end{equation*}
During preliminary numerical calculations it was found that the $\sin (\phi_K-\phi_3)$ asymmetry 
\begin{equation}\label{eq:asymmetry-definition}
    A= \frac{\int \!\! d\phi_K \, [\varsigma \sin(\phi_K-\phi_3)] }{\int \!\!  d\phi_K \, \varsigma} = \frac{\varsigma_1}{2 \varsigma_0}
\end{equation} 
of the distribution is uniquely sensitive to the phase of FFs. It appears only if the relative FFs phase is non-zero.
This is the reason that in the next sections we concentrate on the calculation of asymmetry $\propto \sin(\phi_K-\phi_3)$. Although asymmetry $\propto \cos (\phi_K-\phi_3) $ exists and depends on FFs, it is not uniquely defined by only FF phase. There are other contributions not related to the phase.

\subsection{Hadron current in paraxial approximation}

We are going to calculate components of currents using the exact form of spinors, given in Appendix \ref{secA:twisted_fermion}, Eq.~\eqref{eq:general_spinor:inA}.
We separate the current into two categories: when helicities of $p$ and $\bar{p}$ are the same $(RR/LL)$, and when they are opposite ($RL/LR$).

To simplify analytical calculations, we will use the paraxial approximation:
when cone angle of proton $\theta_1$ is close $0$ and antiproton $\theta_2$ is close to $\pi$.
For completeness, we give full expressions in Appendix \ref{secA:currents} and here show only the result of this approximation. For the $(RR/LL)$ case we get:
\begin{equation}\label{eq:hadron_current_RRLL}
\begin{aligned}
	(RR/LL) & \quad \lambda_1=\lambda_2 \equiv \lambda 
	\\
	\bar{v}_{\lambda} \gamma^0 u_{\lambda} =& 
	2\lambda V_- e^{-i \lambda \phi_{-}}, 
	\\
	\bar{v}_{\lambda} \gamma^z u_{\lambda} =& 
	- 2\lambda W_- e^{-i \lambda \phi_{-}}, 	
	\\
	\bar{v}_{\lambda} \gamma^x u_{\lambda} =& 2 \lambda W_{-} \Big(
	\frac{\varkappa_2}{2|\vec{k}_2|}e^{-i\lambda \phi_+} - \frac{\varkappa_1}{2|\vec{k}_1|}e^{i\lambda \phi_+}
	\Big),  
	\\
	\bar{v}_{\lambda} \gamma^y u_{\lambda} =& i W_{-} \Big(
	\frac{\varkappa_2}{2|\vec{k}_2|}e^{-i\lambda \phi_+} + \frac{\varkappa_1}{2|\vec{k}_1|}e^{i\lambda \phi_+}
	\Big),  
	\\ 
	\bar{v}_{\lambda} u_{\lambda} =&  
	2 \lambda V_+ e^{-i \lambda \phi_-}, 
\end{aligned}
\end{equation}
where $\phi_\pm=\phi_1 \pm \phi_2$ and $M$ is the nucleon mass. We use the following shorthands:
\begin{equation}
\begin{aligned}
    V_\pm =& \sqrt{E^+_1 E_{2}^-} \pm \sqrt{E^-_1 E^+_2} ,
	\\
    W_\pm =& \sqrt{E^+_1 E^+_2} \pm \sqrt{E^-_1 E^-_2},
    \\
    E^{\pm} =& E \pm M.
\end{aligned}
\end{equation}
Note that ``longitudinal'' components and $\bar{v}u$ in Eq.~\eqref{eq:hadron_current_RRLL} are bigger than ``transverse'', which are suppressed by small cone angles that are proportional to $\varkappa_i/2|\vec{k}_i|$.

For opposite helicities the situation changes:
\begin{align}\label{eq:hadron_current_RLLR}
	(RL/LR) & \quad \lambda_1=-\lambda_2 \equiv \lambda \nonumber
	\\
	\bar{v}_{\lambda} \gamma^0 u_{\lambda} =&-V_+ \Big(\frac{\varkappa_1}{2|\vec{k}_1|}e^{i\lambda \phi_-} + \frac{\varkappa_2}{2|\vec{k}_2|}e^{-i\lambda \phi_-} \Big), \nonumber
	\\
	\bar{v}_{\lambda} \gamma^z u_{\lambda} =& 	W_+ \Big(\frac{\varkappa_1}{2|\vec{k}_1|}e^{i\lambda \phi_-} - \frac{\varkappa_2}{2|\vec{k}_2|}e^{-i\lambda \phi_-} \Big), 
	\\
	\bar{v}_{\lambda} \gamma^x u_{\lambda} =& -W_+ e^{-i \lambda \phi_{+}}, \nonumber
	\\
	\bar{v}_{\lambda} \gamma^y u_{\lambda} =& -2 \lambda i W_+ e^{-i \lambda \phi_{+}}, \nonumber
	\\ 
	\bar{v}_{\lambda} u_{\lambda} =&  - V_- \Big(\frac{\varkappa_1}{2|\vec{k}_1|}e^{i\lambda \phi_-} + \frac{\varkappa_2}{2|\vec{k}_2|}e^{-i\lambda \phi_-} \Big). \nonumber
\end{align}
Here ``longitudinal'' and $\bar{v}u$ components are suppressed relatively to transverse.
Moreover, the additional suppression comes from the fact that $V_- \rightarrow 0$ if energies of proton and antiproton are equal.
Therefore, in $RL/LR$ case, $\gamma$ $G_M$ part of the current Eq.~\eqref{eq:hadron_current} gives the biggest contribution to the amplitude, while the part $\propto P^\mu$ is negligible. 
This indicates that in the paraxial limit $RL/LR$ amplitudes are not sensitive to the FF phase.

One can note a pattern in Eqs.~\eqref{eq:hadron_current_RRLL} and \eqref{eq:hadron_current_RLLR}, that longitudinal components are functions of $\phi_-$, while transverse are functions of $\phi_+$. This observation motivates us to separate the scattering amplitude into ``longitudinal'' and ``transverse'' parts, as will be shown in the next section.

\subsection{Lepton current}

In the massless limit, the lepton current is non-zero only when helicities of electron and positron are opposite, $\xi \equiv \xi_3=-\xi_4$. In this case,  current components are
\begin{equation}\label{eq:lepton-current-full}
\begin{aligned}
	L_0 =& -2 \sqrt{E_3 E_4} (c_3 c_4 e^{i\xi \bar{\phi}_-} + s_3 s_4 e^{-i\xi \bar{\phi}_-}), 
	\\
	L_z =& -2 \sqrt{E_3 E_4} (c_3 c_4 e^{i\xi \bar{\phi}_-} - s_3 s_4 e^{-i\xi \bar{\phi}_-}), 
	\\
	L_x =& -2 \sqrt{E_3 E_4} (c_3 s_4 e^{i\xi \bar{\phi}_+} + c_4 s_3 e^{-i\xi \bar{\phi}_+}), 
	\\
	L_y =& 2 i (2\xi) \sqrt{E_3 E_4} (c_3 s_4 e^{i\xi \bar{\phi}_+} - c_4 s_3 e^{-i\xi\bar{\phi}_+}),
\end{aligned}
\end{equation}
where $\bphi_{\pm} \equiv \phi_3 \pm \phi_4$. 
The bar above $\bar{\phi}$ indicates that it is a combination of azimuthal angles of final $e^+e^-$ pair, in contrast to $\phi_{\pm}$ in the hadron current components in Eqs.~\eqref{eq:hadron_current_RRLL} and \eqref{eq:hadron_current_RLLR}.

\begin{figure}
	\centering
	\includegraphics[width=0.7\linewidth]{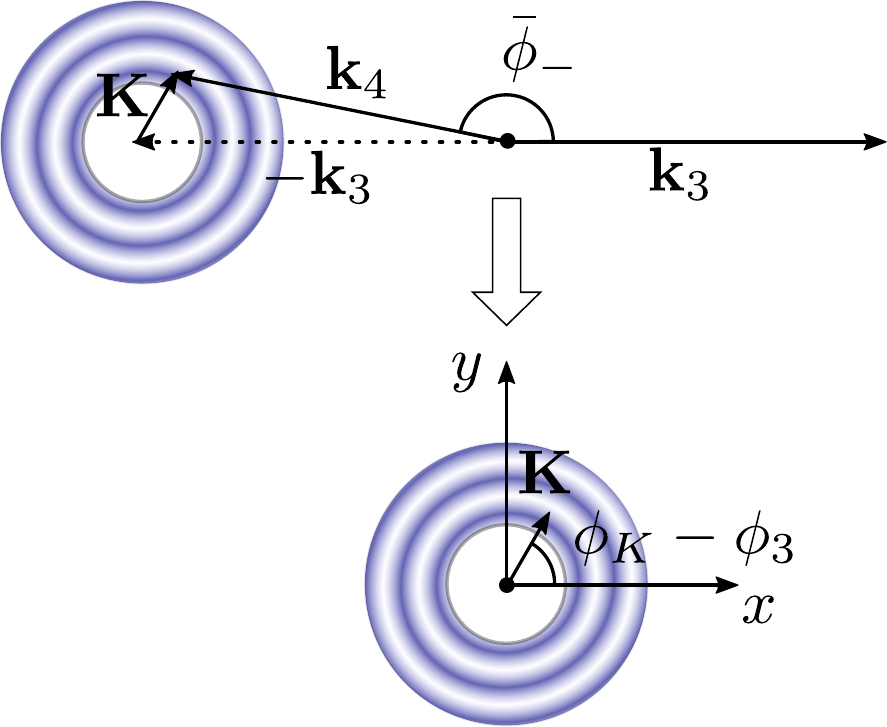}
	\caption{The kinematics of the final $e^+ e^-$ pair in the transverse momentum plane. For a given electron momentum $\bk_3$, the positron momentum $\bk_4=\bK-\bk_3$ has ``freedom'' since $|\varkappa_1-\varkappa_2| \le \bK \le \varkappa_1+\varkappa_2$.}
	\label{fig:ee_kinematics}
\end{figure}

If one keeps lepton current in such a general form, its analytical convolution with the hadron current is verbose. We can simplify calculations by choosing  appropriate particular kinematics, which still pick up important effects, but simplify expressions.
We are going to use the case when 
\begin{equation}\label{eq:lepton-kinematics-choice}
	\varkappa_{1,2},K \ll \varkappa_{3,4}.
\end{equation}
The motivation for such choice is demonstrated in Fig.~\ref{fig:ee_kinematics}.
We will call a plane, spanned by the vortex state axis and the electron momentum $k_3$ as the scattering plane. 
In PW collision, $k_4$ would lay down exactly in the same plane. 
However, in twisted annihilation, it can be off the plane.

With the choice Eq.~\eqref{eq:lepton-kinematics-choice}, the $e^+e^-$ pair is represented by two long, almost opposite vectors in the transverse momentum plane. Their total sum is $\bk_{3}+\bk_{4}=\bK$.
The angle between $\bk_{3}$ and $\bk_{4}$ is close to $\pi$, however we do not want to put it exactly $\pi$, because, as we will see further, the asymmetry in respect to the scattering plane is associated with the relative FF phase.

Essentially, we want to pickup small variations in cross section caused by mirror reflection from the scattering plane, while disregard other minor effects.
One way to do it is to notice that the polar angle $\theta_4 \approx \pi - \theta_3$ varies only slightly if $\bK \ll \bk_{3,4}$, and it symmetric under reflection from the scattering plane.
In contrast, although $\bphi_- \approx - \pi$, it is odd with respect to the reflection. Therefore, we want to keep full azimuthal dependence in exponents $e^{i\xi \bphi_-}$.
With help of these observations, under the condition Eq.~\eqref{eq:lepton-kinematics-choice}, the lepton current Eq.~\eqref{eq:lepton-current-full} simplifies, because $\theta_4 \approx \pi - \theta_3$ leads to $c_4 = s_3$ and $s_4 = c_3$:
\begin{align}\label{eq:lepton_current_simplified}
	L_0 	&= -2\sqrt{E_3 E_4}  \sin\theta_3 \cos (\xi \bar{\phi}_-), \nonumber \\
	L_z 	&= -2 i \sqrt{E_3 E_4} \sin\theta_3 \sin (\xi \bar{\phi}_-),\\
	L_x &=   -2\sqrt{E_3 E_4} (c_3^2 e^{i\xi\bar{\phi}_+} + s_3^2 e^{-i\xi\bar{\phi}_+}), \nonumber\\
	L_y &= 2i 2\xi \sqrt{E_3 E_4} (c_3^2 e^{i\xi\bar{\phi}_+} - s_3^2 e^{-i\xi\bar{\phi}_+}). \nonumber 
\end{align}
Note that, similar to the hadron current, the longitudinal components depend on $\bphi_-$, while the transverse components depend on $\bphi_+$.

\section{Twisted scattering amplitude}

As mentioned in the previous section, due to the different azimuthal dependence of the current components, it is useful to separate the amplitudes into ``longitudinal'' and ``transverse'' parts. The first is formed by time and $z$ components of the currents, while the second is formed by $x$ and $y$ components.
\begin{equation}
\begin{aligned}
    \mathcal{M} &= \mathcal{M}_\parallel + \mathcal{M}_\perp, \\
    \mathcal{M}_{\parallel} &= J_0 L_0 - J_z L_z, \\
    \mathcal{M}_{\perp} &= -J_x L_x - J_y L_y .
\end{aligned}
\end{equation}
Twisted amplitudes, in their turn, can also be represented as the sum of ``longitudinal'' and ``transverse'' parts, since the PW amplitudes $\cM_{a,b}$ enter the equation for the twisted amplitude $\mathcal{J}$ linearly: 
\begin{align}\label{eq:tw_amplitude_def_with_m}
	\mathcal{J} =& \frac{e^{i(m_1-m_2)\phi_K} }{\sin (\delta_1+ \delta_2)} \nonumber
	\\
	&\times \Big(
	\mathcal{M}_a e^{i(m_1\delta_1+m_2\delta_2)}	
    + 
	\mathcal{M}_b e^{-i(m_1\delta_1+m_2\delta_2)}
	\Big).
\end{align}
This equation differs from Eq.~\eqref{eq:tw_amplitude_def_2}, which was written for scalars, by replacing OAM $l_i$ with half-integer total angular momentum $m_i$, since we are now dealing with fermions. 
The double-vortex cross section is proportional to the following sum:
\begin{align}
    |\mathcal{J}|^2 = |\mathcal{J}_{\parallel} + \mathcal{J}_{\perp}|^2 = |\mathcal{J}_{\parallel}|^2 + 2 \Re \{ \mathcal{J}_{\parallel}\mathcal{J}_{\perp}^{*} \} + |\mathcal{J}_{\perp}|^2 .
\end{align}
Every twisted amplitude will be calculated by substituting the corresponding PW amplitude into Eq.~\eqref{eq:tw_amplitude_def_with_m}, as described in Section \ref{section:intro_twisted_states} and Appendix \ref{secA:twScattering}.

\subsection{RR/LL case}
We begin by analyzing the case when the helicities of the proton and antiproton are equal $\lambda_1=\lambda_2 \equiv \lambda$.
Details of the calculation are given in Appendix \ref{secA:RRLLcase}.
In this combination of helicities, the main contribution to the cross section comes from the ``longitudinal'' components of currents:
\begin{align}\label{eq:J2RRLLparallel}
	\sum_\xi|\cJ_{\parallel}|^2 &\propto
32 E_3 E_4 \sin^2 \theta_3 (\mathfrak{C}^{m_1-\lambda}_{m_2-\lambda})^2 |G_M|^2 \nonumber 
\\
           &\times |W_- - \frac{1-G_E/G_M}{2M(1-\tau)} V_+ P_{z}|^2,
\\ 
\mathfrak{C}^{m_1}_{m_2} &= \cos(m_1\delta_1+m_2 \delta_2) .
\end{align}
A common factor $(\frac{4 \pi \alpha}{\sin(\delta_1+\delta_2)s})^2$ is omitted here. 
It is not essential for the discussion, since in twisted scattering we are interested in the oscillations of the cross section, encoded here by $\mathfrak{C}^{m_1-\lambda}_{m_2-\lambda}$, rather than the absolute value of the cross section. 

The Eq.~\eqref{eq:J2RRLLparallel} has the same scattering angle dependence $\sin^2 \theta_3$ as the cross section for head-on annihilation of massive fermions with the same helicities into a pair of massless fermions.
We remind, that we use Eq.~\eqref{eq:lepton-kinematics-choice} to obtain Eq.~\ref{eq:J2RRLLparallel} and it is not valid at near $0$ (or $\pi$) scattering angles.

In the paraxial limit, the interference between ``longitudinal and ``transverse'' twisted amplitudes is suppressed, $\cJ_{\parallel} \cJ_{\perp}^{*} \ll |\cJ_{||}|^2$. However, this interference contains a term $\propto \sin(\phi_K-\phi_3)$, which generates asymmetry and depends on the FFs phase.
It is sum of two parts:
\begin{align}\label{eq:interfJJ_P_RR_LL}
\sum_\xi \Re \{\cJ_{\parallel} \cJ_{P,\perp}^{*} \}
\propto & 
-16 E_3 E_4 \sin 2\theta_3 |G_M|^2 \sin(\phi_k-\phi_3) \nonumber
\\
&\times \varkappa_1  \mathfrak{C}^{m_1-\lambda}_{m_2-\lambda}  (\mathfrak{C}^{m_1-\lambda+1}_{m_2-\lambda} - \mathfrak{C}^{m_1-\lambda-1}_{m_2-\lambda}) \nonumber
\\
&\times \Im \{G_E/G_M\} \frac{(|\vec{k}_1|+|\vec{k}_2|)}{1-\tau}  ,
\end{align}
\begin{align}\label{eq:interfJJ_gamma_RR_LL}
	\sum_\xi \Re \{\cJ_{\parallel} \cJ_{\gamma,\perp}^{*} \}
	\propto & 
   - 16 E_3 E_4 \sin(2\theta_3) |G_M|^2 \sin(\phi_k-\phi_3) \nonumber
	\\
	&\times \lambda P_z \mathfrak{C}^{m_1-\lambda}_{m_2-\lambda} \Big( \frac{\varkappa_1}{|\vec{k}_1|} \mathfrak{C}^{m_1+\lambda}_{m_2-\lambda} +\frac{\varkappa_2}{|\vec{k}_2|} \mathfrak{C}^{m_1-\lambda}_{m_2+\lambda}  \Big) \nonumber
	\\
&\times \Im \{G_E/G_M\} \frac{(|\vec{k}_1|+|\vec{k}_2|)}{1-\tau},
\end{align}
where $\cJ_{\gamma,\perp}$ and $\cJ_{P,\perp}$ are ``transverse'' twisted amplitudes, raised respectively from the $\gamma^\mu$ and $P^\mu$ terms of the hadron current in Eq.~\eqref{eq:hadron_current}.
The sum of Eqs.~\eqref{eq:interfJJ_P_RR_LL} and \eqref{eq:interfJJ_gamma_RR_LL}  simplifies when $K_z=0$:
\begin{align}\label{eq:interfJJ_RRLL}
	\sum_\xi \Re \{\cJ_{\parallel} \cJ_{\perp}^{*} \}
	\propto &  
	16 E_3 E_4 \sin(2\theta_3) |G_M|^2 \sin(\phi_K-\phi_3)  \nonumber
	\\
	&\times 2 \lambda |\mathbf{K}| \Big( \mathfrak{C}^{m_1-\lambda}_{m_2-\lambda} \Big)^2 
	\\
    &\times \Im \{G_E/G_M\} \frac{(|\vec{k}_1|+|\vec{k}_2|)}{1-\tau}. \nonumber
\end{align}
Even though this expression changes sign with helicity, $\mathfrak{C}^{m_1-\lambda}_{m_2-\lambda}$ also differs,  resulting in the asymmetry persisting after summation over helicity $\lambda$ in the unpolarized cross section.

The remained term $|\mathcal{J}_\perp|^2$ is even more suppressed in the paraxial limit than the interference $\cJ_\parallel \cJ_\perp^{*}$. Therefore, we omit it.

\subsection{RL/LR case}

When the proton and antiproton helicities are opposite, $\lambda_{1}=-\lambda_{2}\equiv \lambda$, the situation reverses. In the paraxial limit, the ``longitudinal'' components of the hadron current are suppressed by the small cone opening angle in comparison with ``transverse'' components, Eq.~\eqref{eq:hadron_current_RLLR}. However,  the entire $P^\mu$ part of the current Eq.~\eqref{eq:hadron_current} has additional suppression by $V_-$ (see $\bar{v}_{\lambda}u_{\lambda}$ in Eq.~\eqref{eq:hadron_current_RLLR}).
In the considered kinematics, when $E_1\approx E_2$, it holds that  $V_-\ll V_+,W_{\pm}$. 
Consequently, the contribution of the part containing the relative FF phase is negligible, and the twisted cross section is primarily defined by ``transverse'' $\gamma_\mu G_M$ term of the current Eq.~\eqref{eq:hadron_current}. For detailed calculations, refer to Appendix \ref{secA:RLLRcase}.

The squared modulus of the twisted amplitude after summation over the polarizations of $e^+e^-$ pair is
\begin{equation}\label{eq:J2RL}
\begin{aligned}
\sum_\xi |\cJ_{(RL/LR)}|^2 &\approx 
\sum_\xi |\cJ_{\gamma,\perp}|^2 
\approx 	\left( \frac{4\pi \alpha}{\sin(\delta_1+\delta_2) s} \right)^2 
\\
&\times 32 |G_M|^2 W_+^2 E_3 E_4 (\mathfrak{C}^{m_1-\lambda}_{m_2+\lambda})^2 (1+\cos^2\theta_3). 
\end{aligned}
\end{equation}
Note that $1+\cos^2\theta_3$ dependency of the $RL/LR$ cross section in the paraxial approximation is similar to the behaviour of the annihilation cross section in head-on PW collision for fermions with opposite helicities.

\section{Results and discussion}
In numerical estimations, we use a parametrization for $|G_{E,M}|$ from~\cite{Tomasi:FFfit}
\begin{equation}
	|G_{E,M}| = \frac{9.7}{(1+s/7.1 \GeV^2)(1-s/0.71\GeV^2)^2}.
\end{equation}
Other parametrizations exist, for example~\cite{Kuraev2012:FFmodel}, but the  choice is insignificant for the present analysis. 
Strictly speaking, $|G_E|=|G_M|$ only at threshold $s=4M^2$, and form factors change with $s$~\cite{Tomasi:FFfit}. In the present work, $|G_E|=|G_M|$ is used as a crude estimation, since this qualitatively does not change the observed effect.
The relative phase between FFs $\eta$ is simply parametrized as
\begin{align}\label{eq:FFphase}
	\frac{G_E}{G_M}  =& \frac{|G_E|}{|G_M|} e^{i \eta}. 
\end{align}
We assume that the phase is constant.

\begin{figure}
	\centering
	\includegraphics[width=0.99\linewidth]{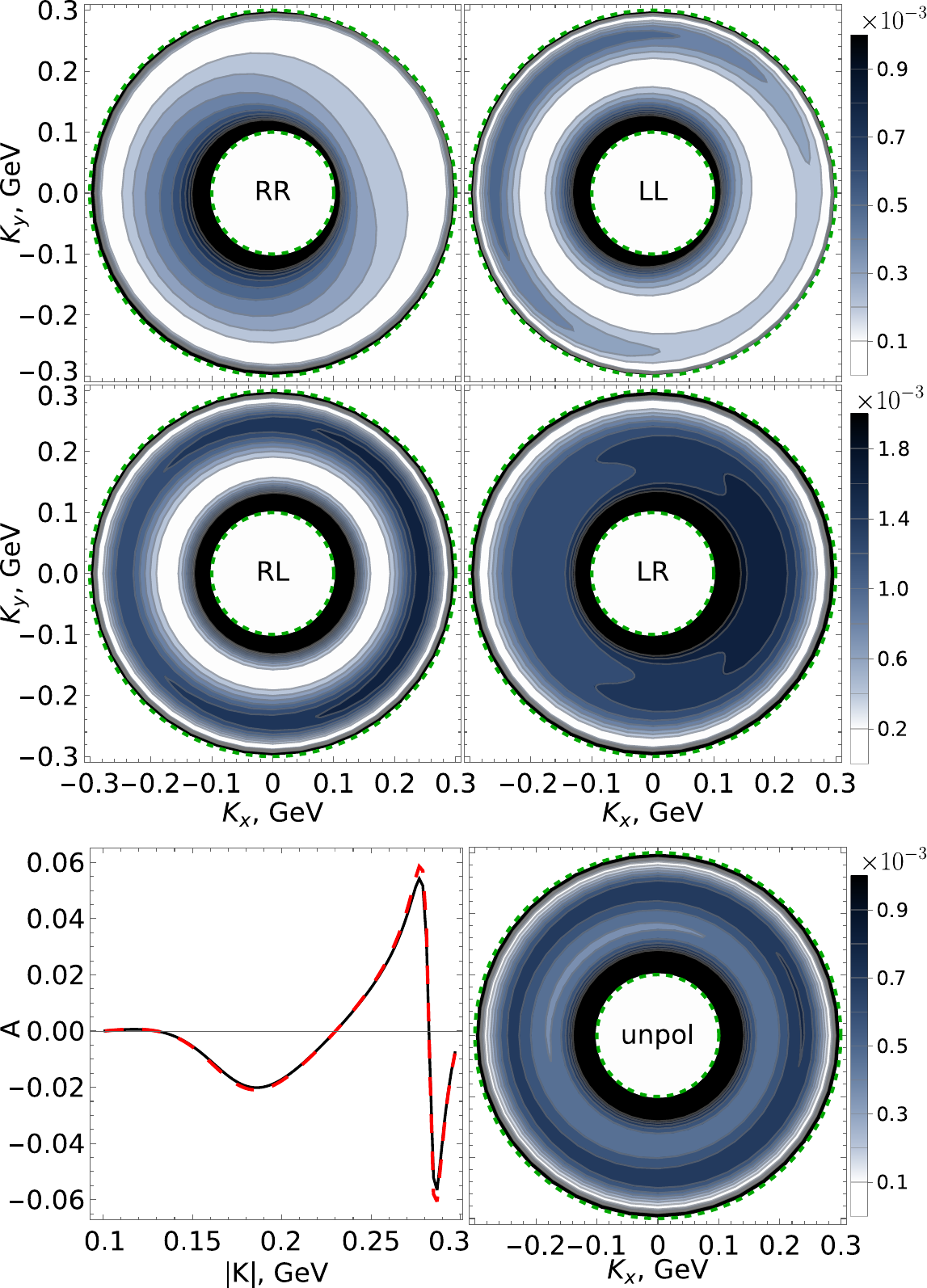}	
	\caption{The cross section distributions over the total transverse momenta $\bK$ for different helicities of the initial $p\bar{p}$ pair: the first row is for the same helicities, the middle row is for opposite helicities. The bottom row shows the asymmetry $A$ calculated as in  Eq.~\eqref{eq:Asin_def} of the unpolarized cross section and the cross section distribution itself. The red dashed line for the asymmetry is the  result of analytical calculation in the paraxial limit.}
	\label{fig:distributions}
\end{figure}

The Fig.~\ref{fig:distributions} demonstrates the result of full numerical calculation of the cross section distributions in total transverse momentum $\bK$ plane for different helicity combinations of the proton and antiproton.
Vortex state parameters and kinematics are chosen as follows:
\begin{align}\label{eq:parameters}
	m_1&=7/2, \quad &m_2 &= 3/2, \nonumber \\
	E_1 &= 1.2\GeV, \quad &E_2&=\sqrt{\varkappa_2^2+M^2+k_{1z}^2}, \nonumber\\
	\varkappa_1 &= 0.2\GeV, \quad &\varkappa_2 &= 0.1\GeV, \\
	\quad K_z &= 0, \quad &\varkappa_3 &= 0.8\GeV, \nonumber
	\\
	\eta &= \pi/2. \quad \nonumber
\end{align}
In such kinematics, the polar scattering angle of the electron  $\theta_3 \approx \pi/4$. We deliberately choose $\varkappa_{1,2}$ to be relatively large in order to see  the difference between the analytical calculation in the paraxial limit and the result without approximation.

A non-zero phase between the FFs modifies the distribution for the $RR$ and $LL$ helicities with respect to the direction of $\bk_{3}$, which is in Fig.~\ref{fig:distributions} is along the positive $\hat{x}$ axis. The non-zero phase slightly rotate the distribution counter-clockwise for $RR$ and clockwise for $LL$.
The $RR$ and $LL$ distributions rotate in opposite directions, and after summation over $p\pbar$ helicity, the asymmetry is preserved due to the different pattern of fringes.
The distributions for the opposite $p\pbar$ helicities $RL/LR$ are symmetrical to the scattering plane and therefore do not contribute to the asymmetry.

\begin{figure}
	\centering
	\includegraphics[width=0.7\linewidth]{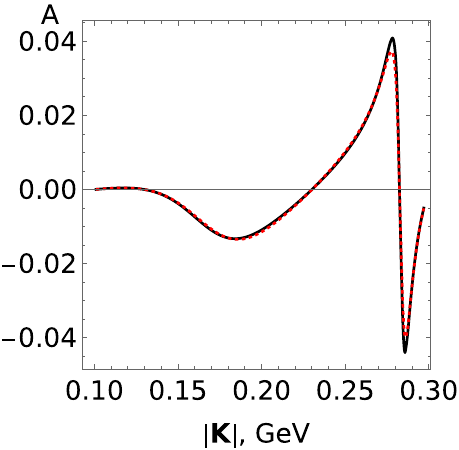}	
	\caption{The asymmetry for the unpolarised cross section distribution integrated over the scattering angle $\theta_3 \in [0,\pi/2]$. The bold line represent the numerical result, while the dotted line corresponds to the analytical result in the paraxial limit.}
	\label{fig:A-integrated}
\end{figure}

To quantify the effect of the FFs phase, we calculate the asymmetry of the cross section distribution in respect to scattering plane, defined in Eq.~\eqref{eq:asymmetry-definition}, as
\begin{equation}\label{eq:Asin_def}
	A = \frac{\int d\phi_K |\mathcal{J}|^2 \sin(\phi_K-\phi_3)}{\int d\phi_K |\mathcal{J}|^2}.
\end{equation}
The result is shown in Fig.~\ref{fig:distributions} in the left panel of the bottom row. 
In the paraxial limit, based on Eq.~\eqref{eq:interfJJ_RRLL} and the FF parametrization Eq.~\eqref{eq:FFphase}, the numerator of Eq.~\eqref{eq:Asin_def} becomes
\begin{multline}\label{eq:interfJJ_RR_LL_integrated}
	\sum_{\xi\lambda} \int d\phi_K \sin(\phi_K-\phi_3) 2 \Re \{\cJ_{\parallel} \cJ_{\perp}^{*} \} \approx 
	\\
	\left( \frac{4\pi \alpha}{\sin(\delta_1+\delta_2) s} \right)^2 
    64 \pi E_3 E_4 \sin(2\theta_3) 
	\\
	\times |\bK| \sin \eta |G_E||G_M| \frac{(|\vec{k}_1|+|\vec{k}_2|)}{1-\tau} \sum_\lambda  \lambda(\mathfrak{C}^{m_1-\lambda}_{m_2-\lambda})^2. 
\end{multline}
The denominator of Eq.~\eqref{eq:Asin_def} in the paraxial limit is the sum of $|\cJ^{(RR/LL)}_{\parallel}|^2$, Eq.~\eqref{eq:J2RRLLparallel}, and $|\cJ^{(RL/LR)}_{\perp}|^2$, Eq.~\eqref{eq:J2RL}.
From Figs.~\ref{fig:distributions} and \ref{fig:A-integrated} one can see that even when cone angles of vortex states are approximately 10 degrees, the paraxial approximation works well.

Notice that in Fig.~\ref{fig:distributions} the total energy $s$ decreases as one moves radially from the inner circle to the outer one, since $s=(E_1+E_2)^2 - \bK^2$. Therefore, in one setup, it is possible to probe different kinematics and $s$ dependency of the FFs phase. It is especially useful if the phase changes rapidly.

The choice of the total angular momentum projection $m_i$ of the vortex states changes the number of interference fringes in the distributions, but it does not affect the absolute value of the asymmetry.  This can be seen from Eq.~\eqref{eq:interfJJ_RRLL}, where the number of fringes is determined by $\mathfrak{C}^{m_1-\lambda}_{m_2-\lambda}$ and the osculation amplitude is proportional to $|\mathbf{K}|$. It is beneficial to keep $|m_{1,2}|$ low, so that distribution has fewer fringes and one does not need to resolve the fine structure of fast oscillations.


When we are talking about twisted states of particles with spin, it is ambiguous when we say ``unpolarized''. 
For a plane-wave state, with its polarization vector independent of spatial coordinates, the unpolarized state is an equal mixture of particles in two orthogonal polarization states, for examples, with $\lambda =+1/2$ and $\lambda =-1/2$.

When we consider a twisted particle with spin $\lambda = \pm 1/2$, we should decide: keep the total angular momentum $m$ fixed, or fix $m - \lambda$.
The answer will depend on the state preparation in specific experimental setup. 
If an experimental device can select states with a single $m$ irrespective of the helicity, then one needs to calculate the process with $|m, \lambda=+1/2\rangle$ and $|m, \lambda=-1/2\rangle$ and perform the averaging. 
If one creates twisted states by imposing a given OAM $l$, then one can describe the unpolarized twisted state as incoherent mixture of particles with
$|m_+ = l+1/2, \lambda =+1/2 \rangle$ and $|m_- = l-1/2, \lambda =-1/2\rangle$.
However, this state will evolve during beam propagation and may become very different in the focal spot due to the spin-orbital interaction.

Therefore, when calculating processes with unpolarized twisted fermion in realistic settings, one must specify according to which definition the twisted beam is unpolarized. 
Fig.~\ref{fig:distributions} shows results when initial proton and antiproton are prepared in state with fixed total angular momentum $m_i$. What will change if incoming particles are prepared with fixed OAM $l_i$?

Working in the paraxial approximation, we can use our previous calculation, replacing $m_i \to l_i + \lambda_i$. 
After changing $m_{1,2}$ in such a way that $l_i \equiv m_i - \lambda_i$ is fixed, from Eq.~\eqref{eq:interfJJ_RR_LL_integrated} follows that in the case of fixed OAM in the paraxial approximation
$A \propto \sum_\lambda \lambda (\mathfrak{C}^{l_1}_{l_2})^2=0$.
We stress out that the distribution still has asymmetry $\propto \cos(\phi_K-\phi_3)$, however it is less useful since contains other contributions, not related with the relative phase of FFs.

Interestingly, the overall shape of the cross section distribution does not depend on the scattering angle $\theta_3$, and we can integrate the cross section over $\theta_3$.
Electron and positron energies $E_3$ and $E_4$ weakly change with $\theta_3$ and $\phi_K$. Therefore, they can be treated as constants when doing analytical integration. Fig.~\ref{fig:A-integrated} shows the asymmetry for the integrated over $\theta_3$ distribution from Fig.~\ref{fig:distributions}. Integration is done in the forward scattering semi-sphere, $0\leq \theta_3\leq\pi/2$, since the asymmetry change sign for $\theta>\pi/2$.
One can see that the overall scale and profile of the asymmetry are preserved.

It is natural to expect, that the first produced vortex states of the proton will be low-energetic.
Therefore, it is interesting to analyse the non-relativistic limit. 
In such a regime 
\begin{equation}
\begin{aligned}
	E&=M+M\beta^2/2, \\
    \vec{k}&=M\vec{\beta}, \\ 
	W_\pm &= M\left(2 + \frac{(|\beta_1| \pm |\beta_2|)^2}{4}\right),
	\\
	V_{\pm} &= M(|\beta_2| \pm |\beta_1|),
\end{aligned}
\end{equation}
where $\beta_i$ is the velocity of the corresponding nucleon.

In the non-relativistic limit, the asymmetry is expressed through velocities as 
\begin{equation}
	\frac{|\bK| (|\vec{k}_1|+|\vec{k}_2|)}{(1-\tau) (W_-^2 + W_+^2)} \approx \frac{1}{4}\frac{\beta_K (|\beta_1|+|\beta_2|) }{\beta_K^2 - |\beta_1|^2 - |\beta_2|^2},
\end{equation}
where $\beta_K$ is the velocity associated with the transverse part of vector $K$.
Therefore, in the non-relativistic regime, the asymmetry is not suppressed by the nucleon mass. It is primarily defined by the cone opening angles of vortex states and becomes small in the paraxial regime. 
Moreover, we stress that the enchantment by $(1-\tau)$ near threshold also implies that the one photon exchange is not a valid approximation for such a  regime.

\begin{figure}
	\centering
	\includegraphics[width=0.99\linewidth]{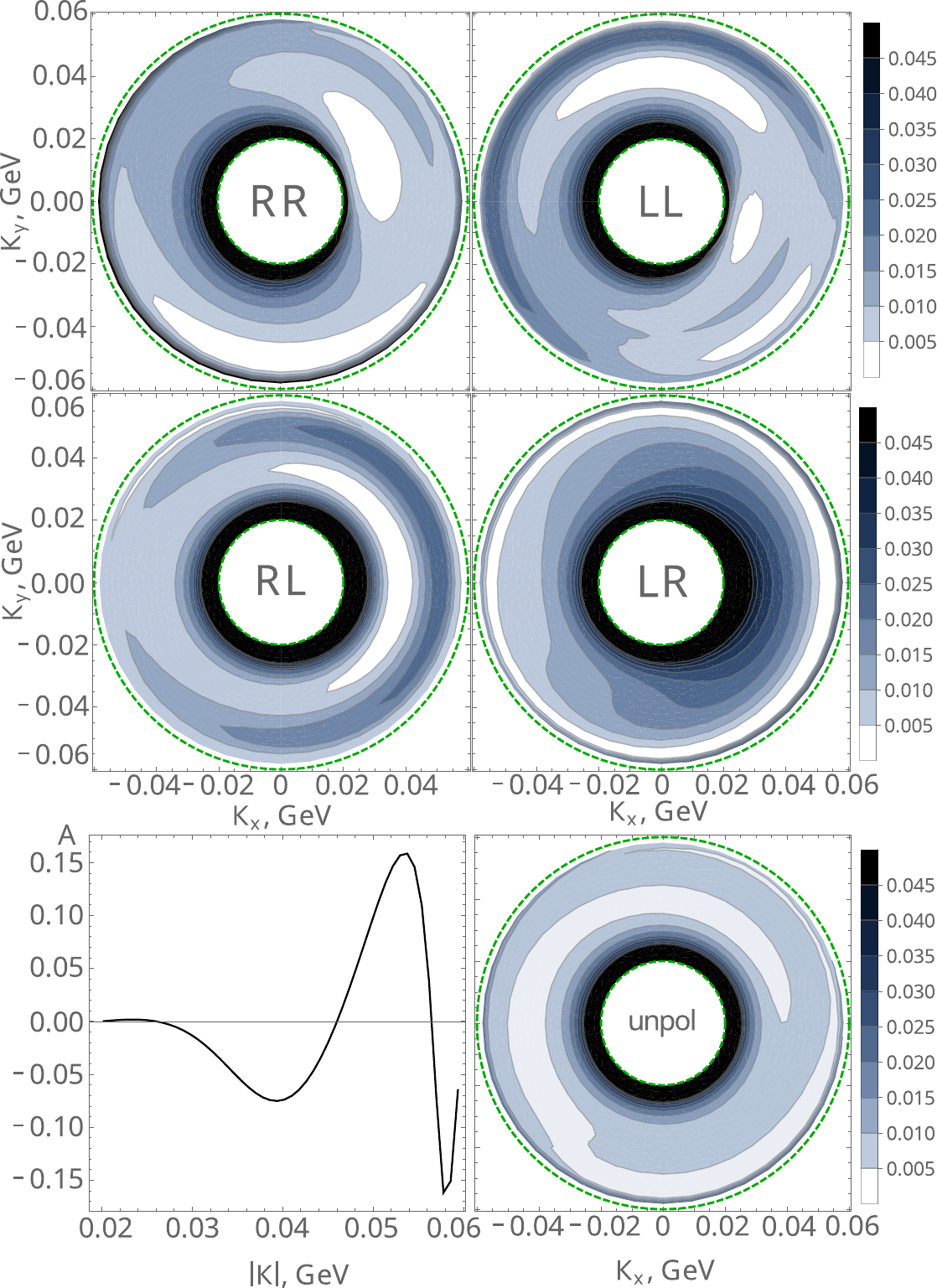}	
	\caption{The cross section distribution over total transverse momentum $\bK$ in non-paraxial non-relativistic kinematics. The notation is similar to Fig.\ref{fig:distributions}. Here the cross section is integrated over the scattering angle $\theta_3$.}
	\label{fig:distributions-NR}
\end{figure}

For demonstration, Fig. \ref{fig:distributions-NR} depicts that the asymmetry can be substantial in the non-relativistic and non-paraxial regime. The kinematics used there are as follows: 
\begin{align}\label{eq:parameters-NR}
	m_1&=7/2, \quad &m_2 &= 3/2, \nonumber \\
	E_1 &= 939\MeV, \quad &E_2&=\sqrt{\varkappa_2^2+M^2+k_{1z}^2}, \nonumber\\
	\varkappa_1 &= 40\MeV, \quad &\varkappa_2 &= 20\MeV, \\
	\quad K_z &= 0, \quad &	\eta &= \pi/2, \quad \nonumber
\end{align} 
The cross section distributions shown in Fig.\ref{fig:distributions-NR} are integrated over $\theta_3$. One can see that the asymmetry can reach 15\%. Notice that $RL$ and $LR$ helicities in non-paraxial regime contribute to asymmetry, but the main effect still comes from $RR$ and $LL$.

One can apply the described technique for studying neutrons.
However, in this case, the asymmetry is suppressed as $|G_E^n|/|G_M^n|$.
In space-like region $G_E^n$ is of the order of 0.05, and $G_M^n$ is of the  order of 1.  If we assume it is the same order in the time-like region, asymmetry of $n\bar{n}$ annihilation will be 20 times less than in $p\bar{p}$.

\section{Conclusion}\label{section:conclusion}

Vortex states open new opportunities for studying particles, unavailable in plane wave scattering.
One of the key features of double-vortex scattering is that the final particles have additional degrees of freedom in the phase space, leading to a non-trivial distribution in the total transverse momentum plane.
The pattern of this distribution is defined by the interference between plane wave amplitudes with different kinematics of incoming waves.
This interference leads, in particular, to the possibility to study Coulomb phase\cite{Ivanov2012:phase,Ivanov2016:CoulombPhase}.

This work proposes another application of this interference: if the scattering amplitude is proportional to some vector, different for two interfering  amplitudes, one can detect the relative phase of the two electromagnetic FFs of the proton even in unpolarized scattering.
By studying the cross section distribution of the final $e^+e^-$ pair in vortex $p\bar p$ annihilation, it was found that its asymmetry is proportional to the sine of the relative phase.
Moreover, such twisted annihilation has a peculiar feature that,
in a single experiment with monochromatic proton and antiproton, one can probe the relative form factor phase at different Mandelstam $s$.

\begin{acknowledgments}
The author thanks Igor P. Ivanov, Pengcheng Zhao and Pengming Zhang for fruitful discussions and review of the manuscript.
\end{acknowledgments}

\appendix

\section{Description of twisted fermion}\label{secA:twisted_fermion}

We construct a twisted fermion state in similar fashion as in \cite{Serbo:2015kia,Ivanov:2016jzt}.
First, we consider a plane wave fermion with mass $M$ and the four-momentum
$k^\mu = (E,\, \bk,\, k_z)$,
where $\bk = |\bk|(\cos\phi_k,\,\sin\phi_k)$, $|\bk| = |\vec k| \sin\theta$, $k_z = |\vec k|\cos\theta$,
and helicity $\lambda = \pm 1/2$  (the eigenvalue of the operator of the spin component along the electron momentum direction).
The plane-wave of such fermion is described by
\begin{equation}
	\Psi_{k \lambda}(r)= {N_{\rm PW} \over \sqrt{2E}}\, u_{k \lambda}\,e^{-i\vec{k}\vec{r}}\,.
	\label{PW:inA}
\end{equation}
The bispinors $u_{k\lambda}$ in above equation are 
\begin{equation}\label{eq:general_spinor:inA}
	\begin{split}
		u_{k\lambda} = \begin{pmatrix}
			\sqrt{E^+}\,w^{(\lambda)} \\
			2 \lambda \sqrt{E^-} \, w^{(\lambda)} 
		\end{pmatrix},&\,
		v_{k\lambda} = \begin{pmatrix}
			-\sqrt{E^-}\,w^{(-\lambda)} \\
			2 \lambda \sqrt{E^+} \, w^{(-\lambda)}
		\end{pmatrix},
		\\
		w^{(+1/2)} = \begin{pmatrix}
			c_i\, e^{-i\phi_i/2}\\
			s_i\, e^{i\phi_i/2}
		\end{pmatrix},&
		\,
		w^{(-1/2)} = \begin{pmatrix}
			-s_i\, e^{-i\phi_i/2} \\
			c_i\, e^{i\phi_i/2}
		\end{pmatrix},
	\end{split}
\end{equation}
where $E^{\pm}=E \pm M$, $c_i \equiv \cos(\theta_i/2)$, and $s_i \equiv \sin(\theta_i/2)$.
The bispinors are normalized as 
$\bar u_{k\lambda_1} u_{k \lambda_2}= 2m\, \delta_{\lambda_1, \lambda_2}$.

We use this basis of plane-wave solutions to construct the Bessel vortex state:
\begin{equation}
	\Psi_{\varkappa m k_z \lambda}(r)= {N_{\rm Bes} \over \sqrt{2E}}\, \int\!\! \frac{d^2 \bk}{(2\pi)^2}\,
	a_{\varkappa m}(\bk)\, u_{k \lambda}\,e^{-i \vec{k}\cdot \vec{r}},
\end{equation}
where the Fourier amplitude is
\begin{equation}
	a_{\varkappa m}(\bk)=(-i)^m \,e^{im\phi_k}\,
	\sqrt{\frac{2\pi}{\varkappa}}\,\delta(|\bk|-\varkappa)\,.
\end{equation}
This state possesses the definite longitudinal momentum $k_z$, the definite conical transverse momentum $|\bk| \equiv \varkappa $, the well-defined helicity $\lambda$ as well as the definite value of the total angular momentum $j_z = m$, which must be half-integer.

Since $m \pm \lambda$ is an integer number, all components of the bispinor $a_{\varkappa,m}$ depend on $\phi_k$ as $\exp[i(m \pm \lambda) \phi_k]$ and are well defined.

Note that, although the Bessel vortex fermion possesses a well-defined helicity, it cannot be interpreted as a state with a well-defined spin projection on any fixed axis. 
The helicity operator involves the direction of $\bk$ and it does not correspond to any component of the spin operator.

The OAM and spin are not separately conserved due to the intrinsic spin-orbital interaction within the vortex electron. 
However, in the paraxial approximation, when $\theta_k \ll 1$, this spin-orbital coupling is suppressed, and one deals with two approximately conserved quantum numbers, $s_z \approx \lambda$ and $l= m - \lambda$.


\section{Double-twisted scattering}\label{secA:twScattering}
Here we give details of the amplitude and cross section for scattering of two twisted states.
The final particles in our $2 \to 2$ annihilation process are plane waves with momenta $k_{3,4}$.
The initial proton and antiproton are described as pure Bessel states with the average longitudinal momenta $k_{1z} > 0$ for the proton and $k_{2z} < 0$ for antiproton.
Since the antiproton propagates in the $-z$ direction, its $z$-projection of AM is $-m_2$. 

The $S$-matrix element of the double-Bessel scattering can be written as
\begin{align}\label{eq:2Bes_Smatrix:inA}
	S =& \frac{N_{Bes}^2}{N_{PW}^2} \int \frac{d^2 \bk_1}{(2\pi)^2}
	\frac{d^2 \bk_2}{(2\pi)^2} a_{\varkappa_1,m_1}(\bk_1) a_{\varkappa_2, -m_2}(\bk_2)
	S_{PW} \nonumber
	\\
	=&  \frac{i (2\pi)^4 \delta(E) \delta(k_z)}{\sqrt{16 E_1 E_2 E_3 E_4}}
	N^2_{Bes} N^2_{PW} \frac{(-i)^{m_1-m_2}}{(2\pi)^3\sqrt{\varkappa_1 \varkappa_2}} \cJ ,
\end{align}
where  $\delta(E)=\delta(E_1+E_2-E_3-E_4)$,  $\delta(k_z)=\delta(k_{1z}+ k_{2z} - k_{3z}-k_{4z})$ and $\mathcal{J}$ is the Bessel scattering amplitude, which is defined as
\begin{align}\label{eq:tw_amplitude_def_1:inA}
	\mathcal{J} =& \int d^2 \bk_1 d^2 \bk_2 e^{i(m_1\phi_1-m_2\phi_2)}
	\delta(|\bk_1|-\varkappa_1)\delta(|\bk_2|-\varkappa_2) \nonumber
	\\
	&\times \delta^{(2)}(\bk_1+\bk_2 - \bK) \mathcal{M}(k_1,k_2,k_3,k_4)
\end{align}
To obtain the differential event rate $d\nu$ one has to take square of Eq.~\eqref{eq:2Bes_Smatrix:inA} and regularise the squares of the delta functions as $[\delta(E)\delta(k_z)]^2 = \delta(E)\delta(k_z) T dz/(2\pi)^2$. Dividing $d\nu$ by the appropriately defined flux and using $dk_{3z} dk_{4z}$ to eliminate $\delta(E)\delta(k_z)$, one obtains the differential cross section (see \cite{Ivanov2022review} and references therein for details).

We represent the cross section simply as
\begin{equation}
	d\sigma \propto |\mathcal{J}|^2 d^2\bk_3 d^2\bk_4.
\end{equation}
The dynamics of the scattering process is determined by the vortex amplitude $\mathcal{J}$ defined in \eqref{eq:tw_amplitude_def_1:inA}. It contains four
integrations and four delta functions, so that the integral can be done exactly for any non-singular $\mathcal{M}$. It is non-zero only when the total transverse momentum $\bK = \bk_1+\bk_2$ lies within the annular region defined by the initial conical momenta:
\begin{equation}
	|\varkappa_1 - \varkappa_2|\leq |\bK| < \varkappa_1 + \varkappa_2
\end{equation}

For any value of $\bK$ inside this region, the integral comes only from two points in the entire $(\bk_1; \bk_2)$ space, at which the momenta $\bk_1 = \varkappa_1(\cos \phi_1; \sin \phi_1)$ and $\bk_2 = \varkappa_2(\cos \phi_2; \sin \phi_2)$ sum up to $\bK$, see Fig.~\ref{fig:2momenta_conf}. 
This happens at the following azimuthal angles:
\begin{alignat}{3}
	\text{configuration a:} \quad &  \phi_1 \to \phi_K+\delta_1  \quad& \phi_2 \to \phi_K-\delta_2 
	\\
	\text{configuration b:} \quad&  \phi_1 \to \phi_K-\delta_1 \quad& \phi_2 \to \phi_K+\delta_2
\end{alignat}
where $\delta_1$ and $\delta_2$ are the internal angles of the triangle with the sides $\varkappa_1$, $\varkappa_2$, and $\bK$:
\begin{align}
	\delta_1 =& \arccos \Big( \frac{\varkappa_1^2-\varkappa_2^2+\bK^2}{2 \varkappa_1 |\bK|} \Big), \\
	\delta_2 =& \arccos \Big( \frac{\varkappa_2^2-\varkappa_1^2+\bK^2}{2 \varkappa_2 |\bK|} \Big).
\end{align}

With these definitions, the result for the vortex amplitude $\mathcal{J}$ can be compactly presented as
\begin{align}\label{eq:tw_amplitude_def_2:inA}
	\mathcal{J} =& \frac{e^{i(m_1-m_2)\phi_K} }{\sin (\delta_1+ \delta_2)} \nonumber
	\\
	&\Big(
	\mathcal{M}_a e^{i(m_1\delta_1+m_2\delta_2)}	+ 
	\mathcal{M}_b e^{-i(m_1\delta_1+m_2\delta_2)}
	\Big).
\end{align}
Notice that the plane-wave amplitudes $\mathcal{M}_a$ and $\mathcal{M}_b$ entering this expression are calculated for the two distinct initial momentum configurations, but the same final momenta $k_3$ and $k_4$. Interference of $\cM_a$ and $\cM_b$ is a crucial change with respect to the PW scattering.

\section{Current components}\label{secA:currents}

The hadron current components associated with $\gamma_\mu$ term in  Eq.~\eqref{eq:hadron_current} are:
\begin{align}
	\bar{v}_{\lambda_2} \gamma_0 u_{\lambda_1} =& -\Big(\sqrt{E^+_1 E_{2}^-} - 4\lambda_1\lambda_2 \sqrt{E^-_1 E^+_2}\Big) w_2^{(-\lambda_2)\dagger} w_1^{(\lambda_1)}, 
	\\
	\bar{v}_{\lambda_2} \gamma_i u_{\lambda_1} =& 
	\Big( 
	2 \lambda_2 \sqrt{E^+_1 E^+_2} - 2 \lambda_1 \sqrt{E^-_1 E^-_2}
	\Big) 
	w_2^{(-\lambda_2)\dagger} \sigma_i w_1^{(\lambda_1)}.
\end{align}
The second part, proportional to the momenta $P^{\mu}$, is
\begin{align}
	\bar{v}_{\lambda_2} u_{\lambda_1} &= - 
	\Big(
	\sqrt{E^+_1 E^-_2 } + 4\lambda_1\lambda_2 \sqrt{E^-_1 E^+_2}
	\Big)
	w_2^{(-\lambda_2)\dagger} w_1^{(\lambda_1)},
\end{align}
where $E_i^\pm=E_i\pm M$.

Further, the following notation will be used:
\begin{align}
	\sqrt{E^+_1 E_{2}^-} \pm \sqrt{E^-_1 E^+_2} 
	=& V_\pm ,
	\\
	\sqrt{E^+_1 E^+_2} \pm \sqrt{E^-_1 E^-_2}
 	=& W_\pm.
\end{align}
The combinations of $w$ spinors and Pauli matrices can be calculated using Eq.~\eqref{eq:general_spinor:inA}. We separate results into two cases: when helicities of $p$ and $\bar{p}$ are the same $(RR/LL)$, and when helicities are opposite ($RL/LR$).
\begin{equation}
\begin{aligned}\label{eq:spinor_combinations:inA}
	(RR/LL) & \quad \lambda_1=\lambda_2 = \lambda:& \nonumber 
	\\
	w_2^{(-\lambda)\dagger} w_1^{(\lambda)}
	&= (2\lambda) (s_1 c_2 e^{i\lambda \phi_-} - c_1 s_2 e^{-i\lambda \phi_-}),
	\\
	w_2^{(-\lambda)\dagger} \sigma_x w_1^{(\lambda)} &=  - (s_1 s_2 e^{i \lambda \phi_+} - c_1 c_2 e^{-i \lambda \phi_+}),
	\\
	w_2^{(-\lambda)\dagger} \sigma_y w_1^{(\lambda)} &= i (2\lambda) ( s_1 s_2 e^{i \lambda \phi_+} + c_1 c_2 e^{-i \lambda \phi_+} ),
	\\
	w_2^{(-\lambda)\dagger} \sigma_z w_1^{(\lambda)} &= - 
	(	s_1 c_2 e^{\lambda i \phi_-} + c_1 s_2 e^{-\lambda i \phi_-} 	),
	\\
	(RL/LR)& \quad \lambda_1=-\lambda_2 = \lambda:&   \nonumber
	\\
	  w_2^{(\lambda)\dagger} w_1^{(\lambda)} &=
	  s_1 s_2 e^{i\lambda \phi_-} + c_1 c_2 e^{-i \lambda \phi_-},  
	  \\
	 w_2^{(\lambda)\dagger} \sigma_x w_1^{(\lambda)} &= (2\lambda) (s_1 c_2 e^{\lambda i \phi_+}  + c_1 s_2 e^{-\lambda i \phi_+} )  , 
	  \\
	  w_2^{(\lambda)\dagger} \sigma_y w_1^{(\lambda)} &=  -i ( s_1 c_2 e^{\lambda i \phi_+} - c_1 s_2 e^{-\lambda i \phi_+}) ,
	  \\
	  w_2^{(\lambda)\dagger} \sigma_z w_1^{(\lambda)} &= -2\lambda 
	  (s_1 s_2 e^{\lambda i \phi_-} - c_1 c_2 e^{-\lambda i \phi_-} ), 
\end{aligned}
\end{equation}
where $\phi_\pm=\phi_1 \pm \phi_2$.
Notice that the helicity of an anti-particle is opposite to the helicity of spinor $w$ it is described by.
One can observe a pattern that time and $z$ components are functions of $\phi_-$, while $x$ and $y$ components are functions of $\phi_+$. This observation motivates to separate amplitude onto ``longitudinal'' and ``transverse'' parts.

Analytical calculations simplifies in the paraxial approximation:
when cone angle of proton $\theta_1$ is close to $0$ and antiproton $\theta_2$ is close to $\pi$.
Therefore
\begin{alignat*}{1}
	s_1 s_2 \to& \frac{\theta_1}{2} = \frac{\varkappa_1}{2|\vec{k}_1|}, \nonumber 
	\\
	c_1 c_2 \to& \frac{\pi-\theta_2}{2} = \frac{\varkappa_2}{2|\vec{k}_2|}, \nonumber 
	\\
	s_1 c_2 \to&  \frac{1}{4}\theta_1(\pi-\theta_2) = \frac{\varkappa_1 \varkappa_2}{4|\vec{k}_1| |\vec{k}_2|},  
	\\
	c_1 s_2 \to& 1.
\end{alignat*}

Applying this approximation, for the $(RR/LL)$ case one gets:
\begin{align}\label{eq:hadron_current_RRLLcomponents:inA}
	(RR/LL) \nonumber
	\\
	\bar{v}_{\lambda} \gamma^0 u_{\lambda} =& 
	2\lambda V_- e^{-i \lambda \phi_{-}}, \nonumber
	\\
	\bar{v}_{\lambda} \gamma^z u_{\lambda} =& 
	- 2\lambda W_- e^{-i \lambda \phi_{-}}, 	
	\\
	\bar{v}_{\lambda} \gamma^x u_{\lambda} =& 2 \lambda W_{-} \Big(
	\frac{\varkappa_2}{2|\vec{k}_2|}e^{-i\lambda \phi_+} - \frac{\varkappa_1}{2|\vec{k}_1|}e^{i\lambda \phi_+}
	\Big), \nonumber 
	\\
	\bar{v}_{\lambda} \gamma^y u_{\lambda} =& i W_{-} \Big(
	\frac{\varkappa_2}{2|\vec{k}_2|}e^{-i\lambda \phi_+} + \frac{\varkappa_1}{2|\vec{k}_1|}e^{i\lambda \phi_+}
	\Big), \nonumber 
	\\ 
	\bar{v}_{\lambda} u_{\lambda} =&  
	2 \lambda V_+ e^{-i \lambda \phi_-}. \nonumber
\end{align}
Note that the ``longitudinal'' and $\bar{v}u$ components dominate in the paraxial limit, while the ``transverse'' components are suppressed by small cone angles, since $\varkappa_i/2|\vec{k}_i| \ll 1$.

In the case of opposite helicities the current components become
\begin{align}
	(RL/LR) \nonumber
	\\
	\bar{v}_{\lambda} \gamma^0 u_{\lambda} =&-V_+ \Big(\frac{\varkappa_1}{2|\vec{k}_1|}e^{i\lambda \phi_-} + \frac{\varkappa_2}{2|\vec{k}_2|}e^{-i\lambda \phi_-} \Big), \nonumber
	\\
	\bar{v}_{\lambda} \gamma^z u_{\lambda} =& 	W_+ \Big(\frac{\varkappa_1}{2|\vec{k}_1|}e^{i\lambda \phi_-} - \frac{\varkappa_2}{2|\vec{k}_2|}e^{-i\lambda \phi_-} \Big), 
	\\
	\bar{v}_{\lambda} \gamma^x u_{\lambda} =& -W_+ e^{-i \lambda \phi_{+}}, \nonumber
	\\
	\bar{v}_{\lambda} \gamma^y u_{\lambda} =& -2 \lambda i W_+ e^{-i \lambda \phi_{+}}, \nonumber
	\\ 
	\bar{v}_{\lambda} u_{\lambda} =&  - V_- \Big(\frac{\varkappa_1}{2|\vec{k}_1|}e^{i\lambda \phi_-} + \frac{\varkappa_2}{2|\vec{k}_2|}e^{-i\lambda \phi_-} \Big). \nonumber
\end{align}
Here the ``longitudinal'' and $\bar{v}u$ components are suppressed relatively to the ``transverse'' components.
Moreover, the $\bar{v}_\lambda u_\lambda$ has an additional suppression from the $V_- \rightarrow 0$ if energies of proton and antiproton are close.
Therefore, contribution of the $P^\mu$ term of the hadron current \eqref{eq:hadron_current} is negligible for $RL/LR$ helicities.

\section{Plane wave and vortex scattering amplitudes}\label{secA:PWamplitudes}

Since the components of currents are divided into two categories based on the azimuthal dependence, the scattering amplitudes follow the same and divided into the ``longitudinal'' and ``transverse'' parts: 
\begin{equation}
\begin{aligned}
\mathcal{M}_{\parallel} &= J_0 L_0 - J_z L_z, \\
\mathcal{M}_{\perp} &= - J_x L_x - J_y L_y .
\end{aligned}
\end{equation}
Then, we need to calculate these plane wave scattering amplitudes for all possible combinations of particles helicities.

\subsection{RR/LL case}\label{secA:RRLLcase}
Using Eqs.~\eqref{eq:hadron_current_RRLLcomponents:inA} and \eqref{eq:lepton_current_simplified}, we obtain the ``longitudinal'' PW scattering amplitude 
\begin{align}\label{eq:pwAmpM_parallel:inA}
	\cM_{\parallel} &= \frac{4\pi\alpha}{s} G_M (2\lambda e^{-i \lambda \phi_{-}})(-2 \sqrt{E_3 E_4}) \sin\theta_3 \nonumber
	\\
	\times & \Big[
	(V_-+Y P_0) \cos(\xi \bar{\phi}_-) + (W_- - Y P_z) i \sin(\xi \bar{\phi}_-) 
	\Big],
\end{align}
where
\begin{align}
	Y =& \frac{1-G_E/G_M}{2M(1-\tau)} V_+.
\end{align}
The $\cos(\xi \bar{\phi}_-)$ term in Eq.~\eqref{eq:pwAmpM_parallel:inA} is suppressed in studied kinematics since $(V_- +Y P_0) \ll (W_{-} - Y P_z)$. In addition, at large scattering angles, when $|\bK | \ll \varkappa_{3,4}$, azimithal angle between electron and positron is $\bar{\phi}_- \approx \pi$.

The ``transverse'' amplitude is the sum of contributions from $\gamma^\mu$ and  $P^\mu$ part of the hadron current Eq.~\eqref{eq:hadron_current}. The amplitude rising from the $\gamma$ term is
\begin{align}\label{eq:PWamp_M_g_perp:inA}
	\mathcal{M}_{a,b}^{(\gamma\perp)} &= \Big(\frac{4\pi\alpha}{s}\Big) G_M W_{-} (2\sqrt{E_3 E_4}) \nonumber
	\\
	&\times \Big(
	\frac{\varkappa_1}{2|\vec{k}_1|}e^{\pm i\lambda \delta_-} \bar{L}_1
	+
	\frac{\varkappa_2}{2|\vec{k}_2|}e^{\mp i\lambda \delta_-} \bar{L}_2
	\Big),
	\\
	\bar{L}_1 &=e^{i\lambda  (2\phi_K - \bar{\phi}_+)} (-2\lambda + 2 \xi \cos\theta_3),
	\\
	\bar{L}_2 &=e^{-i \lambda (2\phi_K - \bar{\phi}_+)} (2\lambda + 2 \xi \cos\theta_3),
\end{align}
where a shorthand $\delta_- \equiv \delta_1-\delta_2$ was used.

The contribution to the ``transverse'' amplitude from $P$ part is
\begin{align}\label{eq:PWamp_M_P_perp:inA}
	\mathcal{M}_{(P,\perp)} &= \Big(\frac{4\pi\alpha}{s}\Big) Y G_M (2\lambda e^{-i \lambda \phi_{-}})(2 \sqrt{E_3 E_4}) \nonumber
	\\
    &\times \Big(\bP_{x} (c_3^2 e^{i \xi \bar{\phi}_+} + s_3^2 e^{-i \xi \bar{\phi}_+}) \nonumber 
    \\ 
    &-2 i \xi \bP_{y} (c_3^2 e^{i \xi \bar{\phi}_+} - s_3^2 e^{-i \xi \bar{\phi}_+}) \Big).
\end{align}
This equation will be simplified later after substitution into Eq.~\eqref{eq:tw_amplitude_def_2:inA}, since the vector $P$ is different for the $a$ and $b$ PW configurations contributing to the twisted amplitude. But first, we calculate the leading ``longitudinal'' twisted amplitude $\cJ_{\parallel}$.

Substitution of Eq.~\eqref{eq:pwAmpM_parallel:inA} into Eq.~\eqref{eq:tw_amplitude_def_2:inA} turns the exponents of azimuthal angles $e^{i\lambda \phi_-}$ to cosines:
\begin{align}\label{eq:Jparallel:inA}
	\mathcal{J}_{\parallel} 
	\approx &  D (16 i \lambda \xi \varkappa_3 G_M)   (W_- - Y P_{z}) \mathfrak{C}^{m_1-\lambda}_{m_2 - \lambda},
	\\
	D &= \frac{e^{i(m_1-m_2)\phi_K}}{\sin(\delta_1+\delta_2)} \frac{4\pi \alpha}{s},
	\\
	\mathfrak{C}^{m_1}_{m_2} &= \cos(m_1\delta_1+m_2 \delta_2).
\end{align}
After summation over polarizations of the $e^+e^-$ pair, the squared modulus of this twisted amplitude is
\begin{align}\label{eq:J2RRLLparallel:inA}
	\sum_\xi|\cJ_{\parallel}|^2=
	|D|^2 32 \varkappa_3^2 (\mathfrak{C}^{m_1-\lambda}_{m_2-\lambda})^2 |G_M|^2 |W_- - Y P_{z}|^2. 
\end{align}
In Eq.~\eqref{eq:Jparallel:inA} we used $\varkappa_3 \approx \sqrt{E_3 E_4} \sin\theta_3$, since $E_3 \approx E_4$.
$|\cJ_{\parallel}|^2$ is the leading contribution to the cross section.

For calculation of the ``transverse'' twisted amplitude, it is useful to express $P^{\mu}$ as
\begin{equation}
	P^\mu = K^\mu - 2 k_1^\mu.
\end{equation}
Then, substituting Eq.~\eqref{eq:PWamp_M_P_perp:inA} into Eq.~\eqref{eq:tw_amplitude_def_2:inA}, one obtains the transverse twisted amplitude, associated with $P$ part of current Eq.~\eqref{eq:hadron_current}
\begin{align}\label{eq:J_P_perp:inA}
	\mathcal{J}_{(P, \perp)} &=  D 4 \lambda Y |G_M| \sqrt{E_3 E_4} \Big( \bK (L_1+L_2) 
	\mathfrak{C}^{m_1-\lambda}_{m_2-\lambda} \nonumber
	\\
	&-2\varkappa_1 
	\Big[ L_1 \mathfrak{C}^{m_1-\lambda+1}_{m_2-\lambda} 
	+ L_2 \mathfrak{C}^{m_1-\lambda-1}_{m_2-\lambda}  \Big] 
	\Big),
\end{align}
where
\begin{align}
	L_1 
	=& (1-2\xi \cos \theta_3) e^{i (\phi_K-\bar{\phi}_+/2)},
	\\
	L_2 
	=& (1+2\xi \cos \theta_3) e^{-i (\phi_K-\bar{\phi}_+/2)}.
\end{align}

Repeating the same procedure for Eq.~\eqref{eq:PWamp_M_g_perp:inA} we obtain
\begin{align}\label{eq:J_g_perp:inA}
	\cJ_{\gamma\perp} =& D G_M (2 \sqrt{E_3 E_4}) W_- \nonumber
	\\
	&\times \Big( \frac{\varkappa_1}{|p_1|} \bar{L}_1 \mathfrak{C}^{m_1+\lambda}_{m_2-\lambda}  + \frac{\varkappa_2}{|p_2|} \bar{L}_2 \mathfrak{C}^{m_1-\lambda}_{m_2+\lambda} \Big) 
\end{align}

The next step is to calculate the interference between the ``longitudinal'' Eq.~\eqref{eq:Jparallel:inA} and ``transverse'' Eqs.~\eqref{eq:J_P_perp:inA},\eqref{eq:J_g_perp:inA} twisted amplitudes. It is the biggest contribution after the leading one Eq.~\eqref{eq:J2RRLLparallel:inA}.
Performing summation over electron-positron polarizations and keeping only terms $\propto \sin(\phi_K-\phi_3)$, two interference contributions are
\begin{align}\label{eq:interfJJ_P_RR_LL:inA}
	\sum_\xi \Re (\cJ_{\parallel} \cJ_{P,\perp}^{*} )
	\approx & |D|^2
	16 E_3 E_4 \sin 2\theta_3 |G_M|^2 W_- \Im (Y) \nonumber
	\\
	&\times \varkappa_1  \mathfrak{C}^{m_1-\lambda}_{m_2-\lambda}  (\mathfrak{C}^{m_1-\lambda+1}_{m_2-\lambda} - \mathfrak{C}^{m_1-\lambda-1}_{m_2-\lambda}) \nonumber
	\\
	&\times \sin(\phi_K-\phi_3),
\end{align}
\begin{align}\label{eq:interfJJ_gamma_RR_LL:inA}
    \sum_{\xi} \Re (\cJ_{\parallel} \cJ_{\gamma,\perp}^{*} )
	\approx & |D|^2 16 E_3 E_4 \sin(2\theta_3) |G_M|^2 W_- \Im(Y)  \nonumber
	\\
	&\times \lambda P_z \mathfrak{C}^{m_1-\lambda}_{m_2-\lambda} \Big( \frac{\varkappa_1}{|\vec{k}_1|} \mathfrak{C}^{m_1+\lambda}_{m_2-\lambda} +\frac{\varkappa_2}{|\vec{k}_2|} \mathfrak{C}^{m_1-\lambda}_{m_2+\lambda}  \Big) \nonumber
	\\
	&\times \sin(\phi_K-\phi_3).
\end{align}

These two terms could be combined together and simplified if $K_z=0$, using
\begin{equation}
\begin{aligned}
		2 \lambda \mathfrak{C}^{m_{1}-\lambda}_{m_{2}+\lambda} =&  2\lambda (\mathfrak{C}^{m_{1}-\lambda}_{m_{2}-\lambda} \mathfrak{C}^0_{2\lambda} - \mathfrak{S}^{m_{1}-\lambda}_{m_{2}-\lambda}\mathfrak{S}^0_{2\lambda}) 
		\\
		=& 2\lambda \mathfrak{C}^{m_{1}-\lambda}_{m_{2}-\lambda} \mathfrak{C}_{1}^0 - \mathfrak{S}^{m_{1}-\lambda}_{m_{2}-\lambda}\mathfrak{S}_{1}^0,
\end{aligned}
\end{equation}
where $\mathfrak{S}_{m_2}^{m_1}\equiv\sin(m_1 \delta_1 + m_2\delta_2)$ and  $\mathfrak{C}^0_{2\lambda}=\cos 2\lambda \delta_2$.
Collecting coefficients in front of $\varkappa_{1,2}$
\begin{align}
	\varkappa_1:\quad & -2\lambda \mathfrak{C}^{m_1+\lambda}_{m_2-\lambda} + \mathfrak{C}^{m_1-\lambda+1}_{m_2-\lambda} - \mathfrak{C}^{m_1-\lambda-1}_{m_2-\lambda} \nonumber
	\\
	&= -2\lambda \mathfrak{C}^{m_1-\lambda}_{m_2-\lambda} \mathfrak{C}^{1}_{0} - \mathfrak{S}^{m_1-\lambda}_{m_2-\lambda} \mathfrak{S}^1_0,
	\\
	\varkappa_2:\quad & -2\lambda \mathfrak{C}^{m_1-\lambda}_{m_2+\lambda} = -2\lambda \mathfrak{C}^{m_1-\lambda}_{m_2-\lambda} \mathfrak{C}^{0}_{1} + \mathfrak{S}^{m_1-\lambda}_{m_2-\lambda} \mathfrak{S}^0_1 ,
\end{align}
and noting from geometrical meaning (see Fig.~\ref{fig:2momenta_conf}) that
\begin{align}
\varkappa_{1} \cos \delta_{1} + \varkappa_{2} \cos \delta_{2} &= |\bK|,
	\\
\varkappa_1 \sin \delta_1 - \varkappa_2 \sin \delta_2 &=0, 
\end{align}
the interference term is simplified to
\begin{align}
    \sum_{\xi \lambda} \Re (\cJ_{\parallel} \cJ_{\perp}^{*} )
	\approx & -|D|^2 16 E_3 E_4 \sin(2\theta_3) W_- \Im(Y) \nonumber
	\\ 
	&\times \sin(\phi_K-\phi_3) \sum_\lambda 2\lambda \bK (\mathfrak{C}^{m_1-\lambda}_{m_2-\lambda} )^2. \nonumber
\end{align}

The square of the ``transverse'' twisted amplitude $|\mathcal{J}_\perp|^2$ is suppressed even more than the interference term  $\cJ_{\parallel} \cJ_{\perp}^{*}$. Therefore, in paraxial limit, it could be neglected.

\subsection{RL/LR case}\label{secA:RLLRcase}

In the case of opposite helicities of proton and antiproton, ``longitudinal'' PW amplitudes $\cM_{\parallel} $ are suppressed by small opening cone angles:
\begin{align}
	\cM_{\gamma,\parallel} =& - \left( \frac{4\pi\alpha}{s}\right) G_M W_+ 4 i \xi \sqrt{E_3 E_4} \sin\theta_3 \nonumber
	\\
	& \times
	\Big( \frac{\varkappa_1}{2|\vec{k}_1|} e^{i\lambda \phi_-} - \frac{\varkappa_2}{2|\vec{k}_2|} e^{-i\lambda \phi_-}\Big),
	\\
	\cM_{P,\parallel} =& \left( \frac{4\pi\alpha}{s}\right) Y_- P_z 4 i \xi \sqrt{E_3 E_4} \sin\theta_3 \nonumber
	\\
	&\times 
	\Big( \frac{\varkappa_1}{2|\vec{k}_1|} e^{i\lambda \phi_-} + \frac{\varkappa_2}{2|\vec{k}_2|} e^{-i\lambda \phi_-}\Big),
\end{align}
where $Y_-= \frac{1-G_E/G_M}{2M(1-\tau)} V_-$.
Substituting these amplitudes into Eq.~\eqref{eq:tw_amplitude_def_2:inA}, we obtain the ``longitudinal'' twisted amplitude
\begin{align}
	\cJ_{\parallel} =&  4 i \xi D G_M \sqrt{E_3 E_4} \sin\theta_3
	\Big( (-W_+ \pm Y_- P_z)\frac{\varkappa_1}{|\vec{k}_1|} \mathfrak{C}^{m_1+\lambda}_{m_2+\lambda} \nonumber
	\\
	&+ (W_+ \pm Y_- P_z) \frac{\varkappa_2}{|\vec{k}_2|} \mathfrak{C}^{m_1-\lambda}_{m_2-\lambda} \Big) 
	\\
	\approx&  4 i \xi D G_M W_+ \varkappa_3
	\Big( 
	\frac{\varkappa_2}{|\vec{k}_2|} \mathfrak{C}^{m_1-\lambda}_{m_2-\lambda} 
	-\frac{\varkappa_1}{|\vec{k}_1|} \mathfrak{C}^{m_1+\lambda}_{m_2+\lambda} 
	\Big), \nonumber
\end{align}
where $Y_- P_z \ll W_+$ was used to get the last line.

The dominant part of the ``transverse'' PW amplitude is
\begin{align}
	\cM_{\gamma,\perp} =& \left( \frac{4\pi\alpha}{s}\right) (-W_+ G_M) 2 \sqrt{E_3 E_4} e^{-i\lambda (\phi_+ - \bar{\phi}_+)} \nonumber
	\\
	&\times (1+2 \lambda 2\xi \cos\theta_3).
\end{align}
Similar to $\cJ_\parallel$, the part of the amplitude, rising from $P$ term in the hadron current Eq.~\eqref{eq:hadron_current}, is suppressed by to $Y_-$.  Therefore, $\cM_{P,\perp} \ll \cM_{\gamma,\perp}$ and whole twisted amplitude is
\begin{align}
	\cJ_{\perp} \approx \cJ_{\gamma,\perp} =&  - D  G_M W_+ 4 \sqrt{E_3 E_4} e^{-i\lambda (2\phi_K - \bar{\phi}_+)} \nonumber
	\\
	&\times \mathfrak{C}^{m_1-\lambda}_{m_2+\lambda} (1+2 \lambda 2\xi \cos\theta_3).
\end{align}
Taking modulus square and summing over polarization of $e^+e^-$ pair one gets
\begin{equation}\label{eq:J2RL:inA}
    \sum_\xi |\cJ_{\gamma,\perp}|^2 = 32 |D G_M|^2 W_+^2 E_3 E_4 (\mathfrak{C}^{m_1-\lambda}_{m_2+\lambda})^2 (1+\cos^2\theta_3). 
\end{equation}

The key observation here is that in the case of $RR/LL$ helicities the interference term $\cJ_\parallel \cJ_{\perp}^*$ contains only $|G_M|^2$, since $Y_-$ part with the electric form factor $G_E$ is suppressed.  It is in contrast with $RR/LL$ case, where the interference term contains both $G_M$ and $G_E$, and is sensitive to their relative phase.


\bibliography{vortex.bib}

\end{document}